\documentclass[12pt]{article}

\usepackage{amssymb}
\usepackage{amstext}
\usepackage{graphicx}

\def\be{\begin{equation}}

\def\ee{\end{equation}}

\def\ba{\begin{eqnarray}}

\def\ea{\end{eqnarray}}

\def \mul {\odot}

\def \ba {\bar}

\def \2 {{1 \over 2}}
\def \3 {{1 \over 3}}
\def \4 {{1 \over 4}}
\def \5 {{1 \over 5}}
\def \6 {{1 \over 6}}
\def \7 {{1 \over 7}}
\def \8 {{1 \over 8}}
\def \9 {{1 \over 9}}
\def \0 { \infty}

\begin{document}
\pagestyle{empty}
\begin{flushright}
\begin{tabular}{ll}
SWAT-03/390 & \\
hep-th/0311254 & \\
November 2003 & \\ [.5in]
\end{tabular}
\end{flushright}
\begin{center}
{\Large {\bf{Massive gauge-invariant field theories on spaces of constant curvature}}} \\ [.7in]
{\large{Paul de Medeiros}} \\ [.3in]
{\emph{Physics Department, University of Wales Swansea, \\ Singleton Park, Swansea SA2 8PP, U.K.}} \\ [.3in]
{\tt{p.de.medeiros@swan.ac.uk}} \\ [.7in]
{\large{\bf{Abstract}}} \\ [.2in]
\end{center}
Gauge fields of mixed symmetry, corresponding to arbitrary representations of the local Lorentz group of the background spacetime, arise as massive modes in compactifications of superstring theories. We describe bosonic gauge field theories on constant curvature spaces whose fields are in irreducible representations of the general linear group corresponding to Young tableaux with two columns. The gauge-invariant actions for such fields are given and generally require the use of auxiliary fields and additional mass-like terms. We examine these theories in various (partially) massless regimes in which each of the mass-like parameters vanishes. We also make some comments about how the structure extends for gauge fields corresponding to arbitrary Young tableaux.    
\clearpage
\pagestyle{plain}
\pagenumbering{arabic}
\setlength{\parskip}{.05in}
{\scriptsize \line(1,0){475}
\tableofcontents
\line(1,0){475}}

\setlength{\parskip}{.15in}


\section{Introduction}

Gauge fields of mixed symmetry type are present in the spectra of massive modes after the compactification of superstring theories. The details of the compactification determines the representation of the local Lorentz group of the spacetime that these massive fields inhabit. In four-dimensional Minkowski space all the massive bosonic fields correspond to totally symmetric tensors, which can be labeled by the integer spin classifying irreducible representations of $SO(3)$. However, in spacetime dimension $D >4$ these irreducible tensor representations generally correspond to Young tableaux with arbitrary numbers of rows and columns.

For the case of massless fields, the description of these more exotic tensor gauge theories in higher dimensional Minkowski space has been given in {\cite{keythesis}}, {\cite{keydeMHul1}}, {\cite{keydeMHul2}} and is discussed in many earlier references {\cite{keyFro}}, {\cite{keylabmor}}, {\cite{keysieg}}, {\cite{keytsul}}, {\cite{keyFraSag}}, {\cite{keyDVHen}}, {\cite{keyBekBou}}. The massive case is discussed in {\cite{keyZin3}}, {\cite{keyCurt}}. Each of these works is concerned with formulating the theory in terms of a finite number of gauge fields. 

In anti-de Sitter space, gauge-invariant actions for interacting totally symmetric tensor gauge fields have been constructed in {\cite{keyVasiliev}}, {\cite{keyvaseq}} for $D \leq 5$. In higher dimensions $D \leq 7$, the associated higher spin symmetry algebras and field equations for these theories are also known {\cite{keysezsun}}. The caveat in these descriptions is that they must be written in terms of an infinite number of fields, plus an infinite number of constraints, which can lead to a system with a finite number of physical degrees of freedom.

A further subtlety is that, unlike in flat space, there is no clear definition of masslessness for fields on spaces of constant non-vanishing curvature. That is, many equivalent properties of massless free field theories in flat space like the absence of explicit mass terms in the Lagrangian, gauge invariance, conformal invariance, saturation of unitarity bounds and  decoupling of compensator fields become independent constraints in curved space. Indeed, unlike in flat space, explicit mass-like terms (i.e. terms that are quadratic in fields but containing no derivatives) are generically present in the Lagrangians of gauge theories in curved space. For the purposes of this paper, and as is common in the literature, we will take the property of masslessness of a gauge field in curved space to be synonymous with the existence of a gauge-invariant description for such a field. More precisely, given a massless gauge-invariant action for a single gauge field in flat space then we define the corresponding field in curved space to be {\emph{massless}} if there exists an action for this field that is  invariant under {\emph{all}} the gauge symmetries of the corresponding theory in flat space (and which reduces to this massless action in the limit of vanishing cosmological constant). Alternatively, we define the corresponding field in curved space to be {\emph{partially massless}} if the action for this field is invariant under some (but not all) of the gauge symmetries of the corresponding massless theory in flat space (and which also reduces to this massless action in the flat space limit). A field not falling into the two categories above will be generally be referred to as {\emph{massive}}. An example of a single field that has a massless description in curved space is the graviton (corresponding to the Young tableau 
\, {\tiny \begin{tabular}{|c|c|}\hline
& \\ \hline
\end{tabular}} \,
of $GL(D,{\mathbb{R}})$). An example of a single field that has only a partially massless description in curved space is the field corresponding to the Young tableau 
\, {\tiny \begin{tabular}{|c|c|}\hline
& \\ \hline
   \\ \cline{1-1}
\end{tabular}} \,
 of $GL(D,{\mathbb{R}})$ (sometimes called the \lq elbow' or \lq hook' field). The existence of only partially massless descriptions of certain more general tensor gauge fields in anti-de Sitter space -- corresponding to \lq non-rectangular' Young tableaux -- is discussed in {\cite{keyMetVas}}. Our results are consistent with the observations in this reference.      

In this paper we attempt to describe general tensor gauge theories on spaces of constant curvature
\footnote{Gauge theories on more exotic background geometries (which also arise in string theory) have been discussed in {\cite{keyBuc}}. In particular, these references consider the consistency of the massive graviton in such backgrounds.}
explicitly in terms of a finite number of gauge fields. This is motivated by the work {\cite{keyZin}} and we generalise the results found there to give a description of all bosonic gauge field theories on constant curvature spaces whose fields are in irreducible representations of the general linear group corresponding to Young tableaux with two columns. In section 2 we review some preliminary results concerning the theory of bi-forms on spaces of constant curvature. This will be the natural framework in which to discuss gauge fields corresponding to Young tableaux with two columns. In section 3 we attempt a minimal extension to curved space of a single massless field in flat space. It will be found that a fully gauge-invariant (or massless) action can be constructed for Young tableaux with both columns of equal length and for Young tableaux with one column of zero length (corresponding $p$-form gauge fields). For Young tableaux with columns of unequal (and non-zero) length, it will be found that only a partially gauge-invariant (or partially massless) action can be constructed, after restricting to a set of one-parameter gauge transformations. In section 4 we introduce additional fields and describe fully gauge-invariant actions for massive fields (corresponding to any Young tableau with two columns) in flat and curved space. The derived field equations for these theories will be found to correspond exactly (in a certain (partially) massless limit) to those found by Metsaev {\cite{keyMet}} for general gauge fields in anti-de Sitter space (in physical gauge). In curved space, many of the features we find, like cosmological constant dependent mass bounds and the existence of partially massless limits, are reminiscent of similar studies in {\cite{keyDes1}}, {\cite{keyZin2}}, {\cite{keyDes2}} (for totally symmetric tensor fields)
\footnote{For example, the authors of {\cite{keyDes2}} consider the field equations for massive totally symmetric tensor gauge fields on spaces of constant curvature. In addition to the usual gauge symmetry of the massless bosonic field equations they also find certain reduced gauge symmetries of the massive field equations in (a causally connected region of) de Sitter space, for particular values of the mass-squared (which are proportional to the cosmological constant). It is these values that they call \lq gauge lines' in the (mass-squared)--(cosmological constant) plane and theories with these masses that they call partially massless.}
 and correspond exactly with the examples in {\cite{keyZin}} (for three different Young tableaux with two columns). In section 5 we conclude by discussing how the structure generalises to classify gauge fields corresponding to Young tableaux with any number of rows and columns. In particular we determine the number of auxiliary fields required to describe the most general massive gauge-invariant action.


\section{Bi-form operators}             
We begin by describing the theory of bi-forms whose structure will naturally arise when discussing gauge fields corresponding to Young tableaux with two columns. Much of this material is based on the bi-form structure of tensor fields over the $D$-dimensional flat Minkowski space ${\mathbb{R}}^{D-1,1}$ (with Minkowski metric $\eta$) given in the earlier works {\cite{keythesis}}, {\cite{keydeMHul1}}, {\cite{keydeMHul2}}, though now we consider tensors over any $D$-dimensional orientable pseudo-Riemannian manifold ${\cal{M}}_D ( \Omega )$ (with metric $g$) of constant curvature proportional to $\Omega$ (such that ${\cal{M}}_D ( 0 ) \cong {\mathbb{R}}^{D-1,1}$ in the $\Omega \rightarrow 0$ limit of zero curvature). A discussion of Young tableaux can be found in {\cite{keyHam}}.

We take the Riemann tensor $R$ of ${\cal{M}}_D ( \Omega )$ to have components
%
%
\begin{equation} 
R^{\sigma}_{\;\; \rho\mu\nu} \;\; =\;\; - 2\, \Omega \, g_{\rho [ \mu} \delta_{\nu ]}^{\sigma}
\label{eq:1}
\end{equation} 
where square brackets denote antisymmetrisation. This convention implies the cosmological constant equals ${(D-1)(D-2) \over 2} \Omega$ so that $\Omega$ positive and negative correspond to the $D$-dimensional de Sitter (${dS}_D$) and anti-de Sitter (${adS}_D$) geometries respectively {\cite{keyWald}}. The canonical Levi-Civita connection $\nabla$ (defined such that $\nabla_{\rho} g_{\mu\nu} =0$) is then used to define covariant derivatives which satisfy
%
%
\begin{equation} 
[ \nabla_{\mu} , \nabla_{\nu} ] V_{\rho} \;\; =\;\; - R^{\sigma}_{\;\; \rho\mu\nu} V_{\sigma}
\label{eq:2}
\end{equation} 
on any one-form $V_{\mu}$.  


\subsection{Bi-forms}

Consider the $GL(D,{\mathbb{R}})$-reducible tensor product space of $p$-forms and $q$-forms $X^{p,q} := \Lambda^p \otimes\Lambda^q$ on ${\cal{M}}_D ( \Omega )$ whose elements are                    
%
%
\begin{equation} 
T \;\; =\;\; \frac{1}{p!q!}T_{\mu_1 ...\mu_p\nu_1 ...\nu_q} dx^{\mu_1}\wedge ...\wedge dx^{\mu_p}\otimes dx^{\nu_1}\wedge ...\wedge dx^{\nu_q}
\label{eq:2.1}
\end{equation} 
where the components $T_{\mu_1 ...\mu_p\nu_1 ...\nu_q} = T_{[ \mu_1 ...\mu_p ][ \nu_1 ...\nu_q ]}$ are totally antisymmetric in each of the $\{ \mu \}$ and $\{ \nu \}$ sets of indices separately. No symmetry properties are assumed between the indices $\mu _i$ and the indices $\nu_j$. The tensor field $T\in X^{p,q}$ is well defined and will be referred to as a {\textit{bi-form}}. This definition of a bi-form is useful since one can employ various constructions from the theory of forms acting on the individual $\Lambda^p$ and $\Lambda^q$ subspaces.  

A generalisation of the exterior wedge product defines the bi-form $T \mul T^{\,\prime} \in X^{p+ p^{\prime} ,q+ q^{\prime}}$, for any $T\in X^{p,q}$ and $T^{\,\prime} \in X^{ p^{\prime} , q^{\prime} }$, by
%
%
\begin{eqnarray}
T \mul T^{\,\prime} &=& \frac{1}{p! \, q! \, p^{\prime} ! \, q^{\prime} !}  
T_{\mu_1 ...\mu_p\nu_1 ...\nu_q}{T^{\,\prime}}_{\mu_{p+1} ...\mu_{p+ p^{\prime}}\nu_{q+1} ...\nu_{q+
q^{\prime}}} dx^{\mu_1}\wedge ...\wedge dx^{\mu_{p+ p^{\prime}}} \nonumber \\ 
&&\hspace*{3.5in} \otimes \, dx^{\nu_1}\wedge ...\wedge dx^{\nu_{q+ q^{\prime}}}  
\label{eq:2.2}
\end{eqnarray}
This definition gives the space $X^* := \sum_{(p,q)} \oplus X^{p,q}$ a ring structure with respect to the $\mul$-product and the natural addition of bi-forms. 

The standard operations on differential forms generalise to bi-forms. There are two exterior derivatives on $X^{p,q}$. The left derivative

%
%
\be 
d \; :\; X^{p,q} \rightarrow X^{p+1,q}
\label{eq:2.3}
\ee
and the right derivative
%
\be
{\tilde{d}} \; :\; X^{p,q} \rightarrow X^{p,q+1}
\label{eq:2.4}
\ee
whose actions on $T$ are given by
%
%
\begin{eqnarray}
dT &=& \frac{1}{p!q!}\nabla_{\mu}T_{\mu_1 ...\mu_p \nu_1 ...\nu_q} dx^{\mu}\wedge dx^{\mu_1}\wedge ...\wedge dx^{\mu_p}\otimes dx^{\nu_1}\wedge ...\wedge dx^{\nu_q} \nonumber \\ 
{\tilde{d}} T &=& \frac{1}{p!q!}\nabla_{\nu}T_{\mu_1 ...\mu_p \nu_1 ...\nu_q} dx^{\mu_1}\wedge ...\wedge dx^{\mu_p}\otimes dx^{\nu}\wedge dx^{\nu_1}\wedge ...\wedge dx^{\nu_q}  
\label{eq:2.5}
\end{eqnarray}
Notice that the action of these two exterior derivatives is metric-dependent. This is to be contrasted with, say, the action of exterior derivation on differential forms on a torsionless Riemann manifold. Since covariant derivatives do not commute on ${\cal{M}}_D ( \Omega )$ (except in flat space where $\Omega =0$) then it is clear that neither of the exterior derivatives above square to zero nor do they commute with each other. The $\Omega$-dependent result of these squaring and commuting operations will be given later after some more bi-form operators have been introduced.

Since ${\cal{M}}_D ( \Omega )$ is orientable then one can also use the invariant volume form ${\sqrt{|g|}} \, \epsilon_{\mu_1 ... \mu_D}$ to construct two inequivalent Hodge duality operations on bi-forms. There is a left dual
%
%
\be
*\; :\; X^{p,q} \rightarrow X^{D-p,q}
\label{eq:2.12}
\ee
and a right dual
%
%
\be
{\tilde{*}}\; :\; X^{p,q} \rightarrow X^{p,D-q}
\label{eq:2.13}
\ee
defined by
%
%
\begin{eqnarray} *T &=& \frac{{\sqrt{|g|}}}{p!(D-p)!q!} T_{\mu_1 ...\mu_p\nu_1 ...\nu_q} 
\epsilon^{\mu_1 ...\mu_p}_{\;\;\;\;\;\;\;\;\;\;\mu_{p+1} ...\mu_D} dx^{\mu_{p+1}}\wedge ...\wedge
dx^{\mu_D}\otimes dx^{\nu_1}\wedge ...\wedge dx^{\nu_q} \nonumber \\
\tilde{*}T &=& \frac{{\sqrt{|g|}}}{p!q!(D-q)!} T_{\mu_1 ...\mu_p\nu_1 ...\nu_q} \epsilon^{\nu_1
...\nu_q}_{\;\;\;\;\;\;\;\;\;\;\nu_{q+1} ...\nu_D} dx^{\mu_1}\wedge ...\wedge dx^{\mu_p}\otimes
dx^{\nu_{q+1}}\wedge ...\wedge dx^{\nu_D} \nonumber \\
\label{eq:2.14}
\end{eqnarray}
where indices are raised using the inverse metric tensor $g^{\mu\nu}$. These definitions imply $*^2 = (-1)^{1+p(D-p)}$, ${\tilde{*}}^2 = (-1)^{1+q(D-q)}$ (since is ${\cal{M}}_D ( \Omega )$ is pseudo-Riemannian) and $*\tilde{*}=\tilde{*}*$.

This allows one to also define two inequivalent \lq adjoint' derivatives
%
%
\be
d^{\dagger} \; := \; {(-1)}^{1+D(p+1)} *d* \; :\; X^{p,q} \rightarrow X^{p-1,q} 
\label{eq:2.15}
\ee
and
%
%
\be 
{\tilde{d}}^\dagger \; := \; {(-1)}^{1+D(q+1)} {\tilde{*}} \, {\tilde{d}} {\tilde{*}} \; :\; X^{p,q} \rightarrow X^{p,q-1}
\label{eq:2.16}
\ee
whose actions on $T$ are given by
%
%
\begin{eqnarray}
d^{\dagger} T &=& \frac{1}{(p-1)!q!}\nabla^{\mu_1} T_{\mu_1 \mu_2 ... \mu_p \nu_1 ... \nu_q} dx^{\mu_2}\wedge ...\wedge dx^{\mu_p} \otimes dx^{\nu_1} \wedge ...\wedge dx^{\nu_q} \nonumber \\ 
{\tilde{d}}^\dagger T &=& \frac{1}{p!(q-1)!} \nabla^{\nu_1} T_{\mu_1 ... \mu_p \nu_1 \nu_2 ... \nu_q} dx^{\mu_1} \wedge ... \wedge dx^{\mu_p} \otimes dx^{\nu_2} \wedge ...\wedge dx^{\nu_q}  
\label{eq:2.17}
\end{eqnarray}
Again, neither of these two adjoint derivatives is nilpotent and they do not commute with each other (except when $\Omega = 0$).

A trace operation
%
%
\be
\tau \; :\; X^{p,q} \rightarrow X^{p-1,q-1}
\label{eq:2.18}
\ee
can be defined by
%
%
\begin{equation} 
\tau T \;\; =\;\; \frac{1}{(p-1)!(q-1)!} g^{\mu _1\nu_1} T_{\mu_1 ... \mu_p \nu_1 ... \nu_q} dx^{\mu_2} \wedge
... \wedge dx^{\mu_p} \otimes dx^{\nu_2} \wedge ... \wedge dx^{\nu_q} 
\label{eq:2.19}
\end{equation}
Consequently, one can define two inequivalent \lq adjoint trace' operations
%
%
\be 
\sigma \; := \; {(-1)}^{1+D(p+1)} * \tau * \; :\; X^{p,q} \rightarrow X^{p+1,q-1}
\label{eq:2.20}
\ee
and
%
%
\be  
{\tilde{\sigma}} \; := \; {(-1)}^{1+D(q+1)} {\tilde{*}} \, \tau {\tilde{*}} \; :\; X^{p,q} \rightarrow X^{p-1,q+1}
\label{eq:2.21}
\ee
so that
%
%
\begin{eqnarray} 
\sigma T &=& \frac{{(-1)}^{p+1}}{p!(q-1)!} T_{[\mu_1 ...\mu_p \nu_1 ]\nu_2 ...\nu_q} dx^{\mu_1}\wedge ...\wedge
dx^{\mu_p}\wedge dx^{\nu_1} \otimes dx^{\nu_2}\wedge ...\wedge dx^{\nu_q} \nonumber \\
{\tilde{\sigma}} T &=& \frac{{(-1)}^{q+1}}{(p-1)!q!}T_{[ \mu_1 | \mu_2 ... \mu_p | \nu_1 ...\nu_q ]} dx^{\mu_2}\wedge
...\wedge dx^{\mu_{p}}\otimes dx^{\nu_1}\wedge ...\wedge dx^{\nu_q} \wedge dx^{\mu_1}
\label{eq:2.22}
\end{eqnarray} 

It is also useful to define a map
%
%
\be 
g\; :\; X^{p,q} \rightarrow X^{p+1,q+1}
\label{eq:2.25}
\ee
by
%
%
\begin{equation} 
g T \;\; =\;\; \frac{1}{p!q!} g_{\mu_1 \nu_1} T_{ \mu_2 ...\mu_{p+1} \nu_2 ...\nu_{q+1}} dx^{\mu_1}\wedge ...\wedge dx^{\mu_p} \wedge dx^{\mu_{p+1}} \otimes dx^{\nu_1}\wedge ...\wedge dx^{\nu_q} \wedge dx^{\nu_{q+1}}
\label{eq:2.26}
\end{equation}
where the action in ({\ref{eq:2.26}}) is identical to the $\mul$-product with the metric tensor $g_{\mu\nu}$, so that $g T \equiv g \mul T$.


\subsection{Some operator identities and an inner product}

Having all the necessary bi-form operators at our disposal, it is now convenient to note some identities which will be useful for the forthcoming discussion. On a general element $T \, \in \, X^{p,q}$
%
%
\begin{eqnarray}
d^2 &=& - \Omega \, g\sigma \nonumber \\
d {\tilde{d}} - {\tilde{d}} d &=& \Omega \, (p-q)\, g \nonumber \\
d d^\dagger + d^\dagger d &=& \nabla^2 - \Omega \left( (p(D-p)+q) 1 - g\tau - {\tilde{\sigma}} \sigma \right) \nonumber \\
d \tau^n + {(-1)}^{n+1} \tau^n d &=& n\, {\tilde{d^\dagger}} \tau^{n-1} \nonumber \\
d^\dagger \tau + \tau d^\dagger &=& 0 \nonumber \\
d g + g d &=& 0 \nonumber \\
d^{\dagger} g^n + {(-1)}^{n+1} g^n d^{\dagger} &=& n\, {\tilde{d}} g^{n-1} \label{eq:2.27} \\
d \sigma + \sigma d &=& 0 \nonumber \\
d {\tilde{\sigma}}^n + {(-1)}^{n+1} {\tilde{\sigma}}^n d &=& -n\, {\tilde{d}} {\tilde{\sigma}}^{n-1} \nonumber \\
\tau g - g \tau &=& (D-p-q) 1 \nonumber \\
\sigma {\tilde{\sigma}} - {\tilde{\sigma}} \sigma &=& (p-q) 1 \nonumber \\
\sigma \tau &=& \tau \sigma \nonumber \\
\sigma g &=& g \sigma \nonumber
\end{eqnarray}
where $\nabla^2 := g^{\mu\nu} \nabla_{\mu} \nabla_{\nu}$. Similar relations to those above hold for operators with tildes (they are obtained from ({\ref{eq:2.27}}) by exchanging tilded with untilded operators and switching $p \leftrightarrow q$). In the bi-form notation, many of these relations look the same as those described in flat space in {\cite{keydeMHul1}}, {\cite{keydeMHul2}}. In particular, of the algebraic and differential operators defined above it is only the form of the differential-differential operator relations that is modified by $\Omega$-dependent factors on ${\cal{M}}_D ( \Omega )$.

Some further useful identities are
%
%
\begin{eqnarray}
d^\dagger \left( \sum_{n=0}^{q} { {(-1)}^n \over (n+1) {(n!)}^2 } \, g^{n+1} \tau^n \, T \right) &=& \sum_{n=0}^{q+1} { {(-1)}^n \over {(n!)}^2 } \, g^{n} \tau^n  \, {\tilde{d}} \, T  \nonumber \\
\nonumber \\
{\tilde{d}}^\dagger d^\dagger \left( \sum_{n=0}^{q} { {(-1)}^n \over (n+1) {(n!)}^2 } \, g^{n+1} \tau^n \, T \right) &=& \sum_{n=0}^{q} { {(-1)}^n \over (n+1) {(n!)}^2 } \, g^{n} \tau^{n+1} \, d {\tilde{d}} \, T  \nonumber \\
\label{eq:2.28} \\
\sum_{n=0}^{q+1} { {(-1)}^n \over {(n!)}^2 } \, g^{n} \tau^n g \, T &=& -(D-p-q-1) \, \sum_{n=0}^{q} { {(-1)}^n \over (n+1) {(n!)}^2 } \, g^{n+1} \tau^n \, T \nonumber \\
\nonumber \\
\sum_{n=0}^{q} { {(-1)}^n \over (n+1) {(n!)}^2 } \, g^n \tau^{n+1} g \, T &=& (D-p-q) \, \sum_{n=0}^{q} { {(-1)}^n \over {(n!)}^2 } \, g^{n} \tau^n \, T  \nonumber 
\end{eqnarray}
acting on any bi-form $T \, \in \, X^{p,q}$. The corresponding identities for the tilded operators follow by exchanging the notion of left and right in the manner described above. Note that the identities ({\ref{eq:2.28}}) do not follow for each term in each of the series above, it is only after summation with the precise coefficients above that these results hold.

We conclude this section by describing an appropriate inner product for bi-forms that will be used extensively in later sections when writing action functionals for bi-form gauge fields. Given two bi-forms $U, V \, \in \, X^{p,q} $ then we define their inner product ${(U,V)}_{p,q}$ on ${\cal{M}}_D ( \Omega )$ by the integral
%
%
\begin{equation} 
{(U,V)}_{p,q} \;\; :=\;\; - \frac{1}{p!q!} \int d^D x \, {\sqrt{|g|}} \; U^{\mu_1 ...\mu_{p} \nu_1 ...\nu_{q}} V_{\mu_1 ...\mu_{p} \nu_1 ...\nu_{q}} 
\label{eq:2.29}
\end{equation}
with indices raised using the inverse metric $g^{\mu\nu}$. This definition implies that
%
%
\begin{equation} 
{(dU,V)}_{p,q} \;\; =\;\; - {(U,{d^\dagger}V)}_{p-1,q} 
\label{eq:2.30}
\end{equation}
for any $U \, \in\, X^{p-1,q}$ and $V \, \in\, X^{p,q}$ (with suitable boundary conditions). A similar result also holds for the tilded derivatives. Furthermore, for any $U \, \in\, X^{p-1,q-1}$, $V \, \in\, X^{p,q}$   
%
%
\begin{equation} 
{(gU,V)}_{p,q} \;\; =\;\; {(U, \tau V)}_{p-1,q-1} 
\label{eq:2.31}
\end{equation}

A consequence of these results and ({\ref{eq:2.28}}) are that the inner products
%
%
\begin{equation} 
{\textsf{S}}_A \;\; := \;\; { \left( A, \sum_{n=0}^{q} { {(-1)}^n \over (n+1) {(n!)}^2 } \, g^{n} \tau^{n+1} \, d {\tilde{d}} \, A \right) }_{p,q}  
\label{eq:2.32}
\end{equation}
and
%
%
\begin{equation} 
{\textsf{M}}_A \;\; := \;\; { \left( A, \sum_{n=0}^{q} { {(-1)}^n \over {(n!)}^2 } \, g^{n} \tau^{n} \, A \right) }_{p,q}  
\label{eq:2.33}
\end{equation}
are each well defined for any $A \, \in\, X^{p,q}$ and that
%
%
\begin{equation} 
\delta {\textsf{S}}_A \;\; = \;\; 2\, { \left( \delta A , \sum_{n=0}^{q} { {(-1)}^n \over (n+1) {(n!)}^2 } \, g^{n} \tau^{n+1} \, d {\tilde{d}} \, A \right) }_{p,q}  
\label{eq:2.34}
\end{equation}
and 
%
%
\begin{equation} 
\delta {\textsf{M}}_A \;\; = \;\; 2\, { \left( \delta A , \sum_{n=0}^{q} { {(-1)}^n \over {(n!)}^2 } \, g^{n} \tau^{n} \, A \right) }_{p,q}  
\label{eq:2.35}
\end{equation}
under any infinitesimal change $\delta A$ in $A$. The reason for the nomenclature of these two inner products will become clear in later sections. 
      

\subsection{Young projection}

The bi-forms described above are reducible tensor representations of $GL(D,{\mathbb{R}})$. However, tensors in representations corresponding to Young tableaux with two columns are irreducible under $GL(D,{\mathbb{R}})$. Decomposing a general bi-form in $X^{p,q}$ into its irreducible components corresponds to the $GL(D,{\mathbb{R}})$ Young decomposition of the tensor product of a $p$-form with a $q$-form. Such a decomposition consists of a direct sum of rank $p+q$ tensors, each of which is irreducible under $GL(D,{\mathbb{R}})$, but only one of which is still totally antisymmetric in each set of $p$ and $q$ indices individually (with no further antisymmetries between the two sets of indices
\footnote{For example, when $p \neq 0$ and $q \neq 0$, this caveat excludes the $(p+q)$-form that is present in the decomposition. This tensor is indeed antisymmetric in each of the sets of $p$ and $q$ indices individually but is also totally antisymmetric in all $p+q$ indices.}
). It is this component that is said to be in the irreducible tensor subspace $X^{[p,q]} \subset X^{p,q}$ and which corresponds to a Young tableau with two columns of lengths $p$ and $q$. We refer to such a tensor as being of type $[p,q]$.  

If a general bi-form $T \, \in \, X^{p,q}$ (with $p \geq q$) is in this irreducible subspace then its components $T_{\mu_1 ...\mu_p\nu_1 ...\nu_q} = T_{[ \mu_1 ...\mu_p ][ \nu_1 ...\nu_q ]}$ satisfy {\cite{keyHam}} $T_{[ \mu_1 ...\mu_p\nu_1 ] ...\nu_q} =0$ (and also $T_{\mu_1 ...[ \mu_p\nu_1 ...\nu_q ]} =0$ if $p=q$ -- which then implies $T_{\mu_1 ...\mu_p\nu_1 ...\nu_p} = T_{\nu_1 ...\nu_p\mu_1 ...\mu_p}$ in this case). In bi-form notation this means that a general element $T \, \in \, X^{p,q}$ (with $p \geq q$) is also in $X^{[p,q]}$ only if $\sigma T =0$ (and ${\tilde{\sigma}} T =0$ if $p=q$). One can obtain a unique element in $X^{[p,q]}$, given a bi-form $T \, \in \, X^{p,q}$ via the action of Young projection ${\cal{Y}}_{[p,q]}$ on $T$ which is given by          
%
%
\begin{equation}
{\cal{Y}}_{[p,q]} \circ T \;\; = \;\; T + \sum_{n=1}^{q} { {(-1)}^n \over \left( \prod_{r=1}^{n} r(p-q+r+1) \right) } {\tilde{\sigma}}^n \sigma^n T   
\label{eq:3.1}
\end{equation}
which indeed satisfies 
%
%
\begin{equation}
\sigma \left( {\cal{Y}}_{[p,q]} \circ T \right) \;\; =\;\; 0 
\label{eq:3.2}
\end{equation}
for $p \geq q$ (using ({\ref{eq:2.27}})) and also
%
%
\begin{equation}
{\tilde{\sigma}} \left( {\cal{Y}}_{[p,q]} \circ T \right) \;\; =\;\; 0 
\label{eq:3.3}
\end{equation}
if $p=q$. It is clear from these results that ${\cal{Y}}_{[p,q]} \circ {\cal{Y}}_{[p,q]} = {\cal{Y}}_{[p,q]}$ on any $T \, \in \, X^{p,q}$ and, from ({\ref{eq:3.1}}), that ${\cal{Y}}_{[p,q]}$ acts as the identity operator on $T$ if $T \, \in \, X^{[p,q]}$.

A final important result is that
%
%
\begin{equation} 
{( {\cal{Y}}_{[p,q]} \circ U ,V)}_{p,q} \;\; =\;\; {(U, {\cal{Y}}_{[p,q]} \circ V)}_{p,q} 
\label{eq:3.3a}
\end{equation}
for any bi-forms $U, V \, \in\, X^{p,q}$. 


\section{Single bi-form gauge fields on ${\cal{M}}_D ( \Omega )$}

In this section we attempt to formulate the (massless) action for a single $[p,q]$ tensor gauge field on ${\cal{M}}_D ( \Omega )$ that preserves all the gauge symmetries of the corresponding massless action in flat space. The procedure involves a straightforward generalisation of the construction for a massless field in flat space {\cite{keythesis}}, {\cite{keydeMHul1}}, {\cite{keydeMHul2}}. It will be found that this is only possible for gauge fields corresponding to Young tableaux with either equal length columns (i.e. for $p=q$) or one zero length column (i.e. for $p \geq 0$ and $q=0$, corresponding to $p$-form gauge fields). For unequal lengths (i.e. for $p>q>0$), the theory can only be made invariant under a restricted set of one-parameter gauge transformations in curved space (and is therefore only partially massless). The resulting field equations for these theories will be compared to those found by Metsaev {\cite{keyMet}} for general single gauge fields in anti-de Sitter space (in physical gauge).


\subsection{Massless bi-form gauge theory in flat space}

We begin by briefly reviewing the construction in flat space (a more detailed discussion can be found in {\cite{keythesis}}, {\cite{keydeMHul1}}, {\cite{keydeMHul2}}). The expressions in section 2 can still be used here by replacing the metric $g$ with the Minkowski metric $\eta$ and by setting $\Omega =0$. Consider a $GL(D,{\mathbb{R}})$-irreducible gauge field $A$ which is a tensor of type $[p,q]$ (with $p \geq q$
\footnote{If $p<q$ then one can still use the expressions for $p>q$ by switching the notion of left and right, such that $p \leftrightarrow q$.}
). Irreducibility implies that $\sigma A =0$ (and ${\tilde{\sigma}} A=0$ if $p=q$). The field strength $F_A$ is the type $[p+1,q+1]$ tensor given by
%
%
\begin{equation}
F_A \;\; =\;\; d {\tilde{d}} \, A 
\label{eq:3.4}
\end{equation}
The ordering of exterior derivatives in $F_A$ is unimportant in flat space. This field strength is also irreducible and therefore satisfies the first Bianchi identity $\sigma F_A =0$ (and ${\tilde{\sigma}} F_A = 0$ if $p=q$). Moreover, since the left and right exterior derivatives are both nilpotent in flat space then $F_A$ also satisfies the second Bianchi identities 
%
%
\begin{equation}
d F_A \;\; =\;\; 0 \quad\quad , \quad\quad {\tilde{d}} F_A \;\; =\;\; 0 
\label{eq:3.5}
\end{equation}
This field strength is invariant under the gauge transformation
%
%
\begin{equation}
\delta A \;\; =\;\; {\cal{Y}}_{[p,q]} \circ \left( d \alpha + {\tilde{d}} {\tilde{\alpha}} \right) 
\label{eq:3.6}
\end{equation}
for bi-form parameters $\alpha \, \in \, X^{p-1,q}$ and ${\tilde{\alpha}} \, \in \, X^{p,q-1}$. Notice that these parameters   do not have to be irreducible themselves for gauge-invariance of $F_A$ though we will take them to be irreducible $[p-1,q]$ and $[p,q-1]$ tensors in the forthcoming discussion as this will be necessary in curved space.

The gauge-invariant action for $A$ is given by 
%
%
\begin{equation}
{\textsf{S}}_A \;\; =\;\; {(A, E_A )}_{p,q}  
\label{eq:3.7}
\end{equation}
where $E_A$ is the generalised Einstein tensor for $A$ given by
%
%
\begin{equation}
E_A \;\; =\;\; \sum_{n=0}^{q} { {(-1)}^n \over (n+1) {(n!)}^2 } \, \eta^{n} \tau^{n+1} \, F_A 
\label{eq:3.8}
\end{equation}
Gauge-invariance of ({\ref{eq:3.7}}) follows from the fact that $d^\dagger E_A =0$ and ${\tilde{d}}^\dagger E_A =0$ identically which in turn follow from the second Bianchi identities ({\ref{eq:3.5}}). The action ({\ref{eq:3.7}}) corresponds to ({\ref{eq:2.32}}) in flat space. ({\ref{eq:2.34}}) then implies that the field equation for $A$ is $E_A =0$ in flat space which reduces to the equation 
%
%
\begin{equation}
\tau F_A \;\; =\;\; 0 
\label{eq:3.8a}
\end{equation}
in $D \neq p+q$. This equation is non-trivial, in the sense that it does not imply that the whole field strength $F_A =0$, in dimension $D \geq p+q+2$ (this detail is explained further in {\cite{keydeMHul1}}). The requirement of non-triviality of this field equation will become a consistency condition when we discuss massive fields. In physical gauge, ({\ref{eq:3.8a}}) reduces to the equations
%
%
\begin{equation}
\square \, A \;\; =\;\; 0 \quad , \quad d^\dagger A \;\; =\;\; 0 \quad , \quad {\tilde{d^\dagger}} A \;\; =\;\; 0 \quad , \quad \tau A \;\; =\;\; 0 
\label{eq:3.8b}
\end{equation}
where $\square := \eta^{\mu\nu} \partial_{\mu} \partial_{\nu}$ is the D' Alembertian operator in flat space. These physical gauge equations are still invariant under gauge transformations provided the usual parameters $\alpha$ and ${\tilde{\alpha}}$ now satisfy similar physical gauge equations to ({\ref{eq:3.8b}}) -- as though they too are physical gauge fields of type $[p-1,q]$ and $[p,q-1]$ respectively. The details of going to this physical gauge for general field theories in flat space are discussed in {\cite{keydeMHul2}}. The procedure involves first using the existing gauge symmetry to impose $d^\dagger A =0$ and ${\tilde{d^\dagger}} A =0$. These two equations and the field equation ({\ref{eq:3.8a}}) are found to still be invariant under a reduced set of gauge transformations. This residual gauge symmetry is then used to set $\tau A =0$ on-shell, and the resulting wave equation $\square \, A =0$ is read off from ({\ref{eq:3.8a}}).       

We conclude the discussion in flat space by examining the precise form of the gauge transformations ({\ref{eq:3.6}}) when we take the parameters $\alpha$ and ${\tilde{\alpha}}$ to be $GL(D,{\mathbb{R}})$-irreducible. In the bi-form notation the expressions we derive will take the same form in both flat and curved space (and are particularly useful when doing calculations in curved space). Since we assume $p \geq q$ then there are two distinct cases to consider: $p>q$ and $p=q$. 

For $p>q$, $\alpha \, \in \, X^{[p-1,q]}$ and ${\tilde{\alpha}} \, \in \, X^{[p,q-1]}$ with $p-1 \geq q$ and $p > q-1$ so that 
%
%
\begin{equation}
\sigma \alpha \;\; =\;\; 0 \quad\quad , \quad\quad \sigma {\tilde{\alpha}} \;\; =\;\; 0 
\label{eq:3.9}
\end{equation}
with ${\tilde{\sigma}} \alpha =0$ if $p-1=q$. Using the results of the previous section then one finds the gauge transformation ({\ref{eq:3.6}}) becomes
%
%
\begin{equation}
\delta A \;\; =\;\; d\alpha + {\tilde{d}} {\tilde{\alpha}} + {1 \over (p-q+2)} {\tilde{\sigma}} d {\tilde{\alpha}} 
\label{eq:3.10}
\end{equation}
The above expression is annihilated by $\sigma$, as expected.

For $p=q$, $\alpha \, \in \, X^{[p-1,p]}$ and ${\tilde{\alpha}} \, \in \, X^{[p,p-1]}$ so that 
%
%
\begin{equation}
{\tilde{\sigma}} \alpha \;\; =\;\; 0 \quad\quad , \quad\quad \sigma {\tilde{\alpha}} \;\; =\;\; 0 
\label{eq:3.11}
\end{equation}
and the gauge transformation can be written 
%
%
\begin{equation}
\delta A \;\; =\;\; {\tilde{d}} ( {\tilde{\alpha}} - \sigma \alpha ) + {1 \over 2} {\tilde{\sigma}} d ( {\tilde{\alpha}} - \sigma \alpha ) + \sum_{n=2}^{q} { {(-1)}^n \over \left( \prod_{r=1}^{n} r(p-q+r+1) \right) } {\tilde{\sigma}}^n \sigma^n d \alpha 
\label{eq:3.12}
\end{equation}
Clearly this transformation is more complicated than the $p>q$ case though the leading terms can all be expressed in terms of the one parameter ${\tilde{\alpha}}^\prime := {\tilde{\alpha}} - \sigma \alpha$. Moreover these leading terms take the same form as those for ${\tilde{\alpha}}$ in the $p>q$ case above if one formally sets $p=q$ in ({\ref{eq:3.10}}). The difference is that now $\sigma {\tilde{\alpha}}^\prime = - \sigma^2 \alpha \neq 0$. One can however obtain a consistent set of one parameter transformations for the $p=q$ case by demanding $\sigma^2 \alpha =0$ (even though $\sigma \alpha \neq 0$). This is equivalent to demanding that $\sigma \alpha$ be irreducible. This means that all the $n \geq 2$ terms in the sum in ({\ref{eq:3.12}}) vanish and the transformation for $p=q$ is just as in ({\ref{eq:3.10}}) for $p>q$ but with parameters $\alpha^\prime =0$ and ${\tilde{\alpha}}^\prime$ (since now $\sigma {\tilde{\alpha}}^\prime =0$). This one-parameter gauge transformation is annihilated by both $\sigma$ and ${\tilde{\sigma}}$, as required. It is these one-parameter transformations for tensor gauge fields with $p=q$ that will be preserved when $\Omega \neq 0$.


\subsection{Minimal extension to bi-form gauge theory on ${\cal{M}}_D ( \Omega )$}

By minimal we mean that no additional fields will be introduced at this stage and we try to extend the results in flat space to describe a single type $[p,q]$ tensor gauge field $A$ (with $p \geq q$) on ${\cal{M}}_D ( \Omega )$. The minimal extension of the gauge transformation ({\ref{eq:3.6}}) is to replace the flat space exterior derivatives with those defined in ({\ref{eq:2.5}}) which are covariant on ${\cal{M}}_D ( \Omega )$. This is the only covariant gauge transformation that can be written for $A$ without the addition of extra gauge parameters. 

The proposed field strength $F_A$ is given by
%
%
\begin{equation}
F_A \;\; =\;\; \left( d {\tilde{d}} - \Omega \, k \, g \right) \, A 
\label{eq:3.13}
\end{equation}
for some constant $k$. This reduces to the corresponding flat space field strength ({\ref{eq:3.4}}) when $\Omega =0$. Since the two exterior derivatives no longer commute then one could choose a different ordering of derivatives in ({\ref{eq:3.13}}) though, using ({\ref{eq:2.27}}), this would just correspond to a redefinition of the constant $k$. The field strength $F_A$ is a $GL(D,{\mathbb{R}})$-irreducible tensor of type $[p+1,q+1]$ since it satisfies the first Bianchi identity  
%
%
\begin{equation}
\sigma F_A \;\; =\;\; \left( d {\tilde{d}} - \Omega \, (k+1) \, g \right) \, \sigma A \;\; =\;\; 0
\label{eq:3.14}
\end{equation}
with ${\tilde{\sigma}} F_A =0$ if ${\tilde{\sigma}} A =0$ when $p=q$. Since each of the exterior derivatives is not nilpotent then $F_A$ does not automatically satisfy the second Bianchi identities since
%
%
\begin{eqnarray}
d F_A &=& \Omega \, g \left( {\tilde{d}} \sigma + (k+1) \, d \right) A \nonumber \\
{\tilde{d}} F_A &=& \Omega \, g \left( d {\tilde{\sigma}} + (k+1-p+q) \, {\tilde{d}} \right) A
\label{eq:3.15}
\end{eqnarray}
This means that $d F_A =0$ if $k=-1$ whilst ${\tilde{d}} F_A =0$ if $p=q$ and $k=-1$. Consequently, the covariantised form of the generalised Einstein tensor in flat space ({\ref{eq:3.8}}), defined as the irreducible type $[p,q]$ tensor 
%
%
\begin{equation}
E_A \;\; =\;\; \sum_{n=0}^{q} { {(-1)}^n \over (n+1) {(n!)}^2 } \, g^{n} \tau^{n+1} \, F_A 
\label{eq:3.15a}
\end{equation}
is no longer automatically conserved on ${\cal{M}}_D ( \Omega )$. In particular, the identities
%
%
\begin{eqnarray}
d^\dagger E_A &=& \sum_{n=0}^{q} { n {(-1)}^n \over (n+1) {(n!)}^2 } \, g^{n-1} \tau^{n+1} \, {\tilde{d}} F_A \nonumber \\
{\tilde{d^\dagger}} E_A &=& \sum_{n=0}^{q} { n {(-1)}^n \over (n+1) {(n!)}^2 } \, g^{n-1} \tau^{n+1} \, d F_A 
\label{eq:3.15b}
\end{eqnarray}
imply that $d^\dagger E_A =0$ if $p=q$ and $k=-1$ whilst ${\tilde{d^\dagger}} E_A =0$ if $k=-1$.

Consider now the covariantised action 
%
%
\begin{equation}
{(A, E_A )}_{p,q} \;\; =\;\; {\textsf{S}}_A - \Omega \, k (D-p-q) \, {\textsf{M}}_A
\label{eq:3.15c}
\end{equation}
on ${\cal{M}}_D ( \Omega )$, in terms of the inner products ({\ref{eq:2.32}}) and ({\ref{eq:2.33}}). Notice that the first part ${\textsf{S}}_A$ is kinetic (that is it contains derivatives), just as in flat space, though the second term ${\textsf{M}}_A$ (containing no derivatives) is like a generalised mass term for $A$. Under any infinitesimal variation $\delta A$, the variation of this action (using ({\ref{eq:2.34}}) and ({\ref{eq:2.35}})) is given by 
%
%
\begin{equation}
\delta {(A, E_A )}_{p,q} \;\; =\;\; 2 \, {( A, \delta E_A )}_{p,q} \;\; =\;\; 2 \, {( \delta A, E_A )}_{p,q} 
\label{eq:3.15d}
\end{equation}
Such an action is therefore invariant under the covariantised gauge transformation $\delta A = {\cal{Y}}_{[p,q]} \circ \left( d \alpha + {\tilde{d}} {\tilde{\alpha}} \right)$ provided that either $\delta F_A =0$ (i.e. $F_A$ is gauge-invariant which implies $\delta E_A =0$) or (using ({\ref{eq:2.30}}) and ({\ref{eq:3.3a}})) that $d^\dagger E_A =0$ and ${\tilde{d^\dagger}} E_A =0$. Following ({\ref{eq:3.15b}}), the latter of these two conditions is only satisfied for both $p=q$ and $k=-1$. This is, of course, assuming that both parameters $\alpha$ and ${\tilde{\alpha}}$ are non-vanishing. If one considers the set of one-parameter gauge transformations in ${\tilde{\alpha}}$ (with $\alpha =0$) then the action ({\ref{eq:3.15c}}) is gauge-invariant provided only $k=-1$ (for any $p \geq q$).       

Considering now the first of the gauge-invariance conditions above then one finds that $\delta F_A =0$ only if
%
%
\begin{eqnarray}
{\tilde{\sigma}} {\tilde{\alpha}} &=& (k-p+q) \, \alpha \nonumber \\
\sigma \alpha &=& k \, {\tilde{\alpha}} 
\label{eq:3.16}
\end{eqnarray}
Taking both the gauge parameters to be irreducible and non-vanishing then the equations ({\ref{eq:3.16}}) have a solution for $p=q$. Since the parameters are irreducible then ${\tilde{\sigma}} \alpha =0$ and $\sigma {\tilde{\alpha}} =0$ so ({\ref{eq:3.16}}) then implies $\sigma^2 \alpha =0$ (and ${\tilde{\sigma}}^2 {\tilde{\alpha}} =0$). Therefore the field strength $F_A$ is invariant under the covariantised form of the one-parameter gauge transformations for gauge fields with $p=q$ that were discussed at the end of the previous subsection. This solution is appealing since it coincides with the conditions for $F_A$ to satisfy both the second Bianchi identities ({\ref{eq:3.15}}) if we also take $k=-1$. It is worth noting that another solution of ({\ref{eq:3.16}}) exists for $p >q$ if one removes the restriction that both parameters must be non-vanishing. That is, if one takes ${\tilde{\alpha}} =0$ then $F_A$ is invariant under any irreducible $\alpha$ transformations (satisfying $\sigma \alpha =0$) provided $k=p-q$
\footnote{This would necessarily be the case for a $p$-form gauge field (with $p>0$ and $q=0$).}
. There is no corresponding $p>q$ solution if one takes $\alpha =0$ since ({\ref{eq:3.16}}) imply ${\tilde{\alpha}} =0$ as well in that case.

In summary, there are two classes of theories whose actions ({\ref{eq:3.15c}}) are gauge-invariant on ${\cal{M}}_D ( \Omega )$. The first corresponds to gauge fields with $p=q$ whose actions are completely gauge-invariant provided $k=-1$. The second corresponds to gauge fields with $p>q$ which have two types of solution, each of which are only invariant under a restricted set of one-parameter gauge transformations. The first type has $p>q$ and $k=-1$ which are invariant under gauge transformations with $\alpha =0$. The second type has $p>q$ and $k=p-q$ which are invariant under gauge transformations with ${\tilde{\alpha}} =0$. Clearly it would be desirable to give a formulation that is invariant under both gauge transformations of fields with $p>q>0$ in curved space, such that it correctly reduces to the corresponding fully gauge-invariant theory in the flat space limit. In fact, such a formulation is not possible without adding extra fields. This construction will be described in the next section.

Consider now the form of the derived field equations for the two classes of solutions above in physical gauge. The general procedure for going to physical gauge in curved space follows that described for fields in flat space in the previous subsection. That is, given a gauge-invariant field equation for $A\, \in \, X^{[p,q]}$, one imposes the covariant equations $d^\dagger A =0$ and ${\tilde{d^\dagger}} A =0$
\footnote{These constraint equations correspond to gauge symmetries of the field equation under transformations with parameters $\alpha$ and ${\tilde{\alpha}}$ respectively. Of course, the $p>q>0$ solutions found above are only invariant under a reduced gauge symmetry and so one cannot obtain both $d^\dagger A =0$ and ${\tilde{d^\dagger}} A =0$ for such fields in the usual way (via a sequence of specific gauge transformations). The analysis here is therefore somewhat formal, but will help us understand the nature of the physical gauge equations of motion for general fields with $p>q>0$ that will be obtained from fully gauge-invariant actions in the next section.}
. Residual gauge symmetries of these constraint equations (for which the gauge parameters themselves satisfy particular Laplace equations) are used to set $\tau A =0$ on-shell. The resulting Laplace equation for $A$ in physical gauge is then read off from the field equation.   

More precisely, given the action ({\ref{eq:3.15c}}) for a gauge field $A\, \in \, X^{[p,q]}$, then the field equation is
%
%
\begin{equation}
\tau  \left( d {\tilde{d}} - \Omega \, k\, g \right) A \;\; =\;\; 0
\label{eq:3.21}
\end{equation}
for $D \neq p+q$. In physical gauge, ({\ref{eq:3.21}}) reduces to the equations
%
%
\begin{eqnarray}
\Delta^{(k)}_{[p,q]} \, A &:=& \left( \, \nabla^2 - \Omega \, ( q(D+1-q) + k (D-p-q) ) \, \right) A \;\; =\;\; 0   \nonumber \\
d^\dagger A &=& 0 \quad , \quad {\tilde{d^\dagger}} A \;\; =\;\; 0 \quad , \quad \tau A \;\; =\;\; 0 
\label{eq:3.22} 
\end{eqnarray}
which describe the solutions above in physical gauge when $k=-1$ (for $p=q$ and $p>q$ when $\alpha =0$) and when $k=p-q$ (for $p>q$ when ${\tilde{\alpha}} =0$). Since it will only be these two values of $k$ that will be of interest in the forthcoming discussion, we define
%
%
\begin{equation}
\Delta_{[p,q]} \;\; :=\;\; \Delta^{(-1)}_{[p,q]} \quad \quad , \quad \quad \Delta^{\prime}_{[p,q]} \;\; :=\;\; \Delta^{(p-q)}_{[p,q]} 
\label{eq:3.23}
\end{equation}

The physical gauge equations ({\ref{eq:3.22}}) are identical to those found in {\cite{keyMet}} for the $p=q$, $k=-1$ solution in anti-de Sitter space (the cosmological constant is normalised such that $\Omega =-1$ in {\cite{keyMet}}). Moreover, the gauge transformation for such fields preserves the equations ({\ref{eq:3.22}}) only if the single parameter ${\tilde{\alpha}}^{\prime}$ satisfies the physical gauge equations $\Delta_{[p,q-1]} \, {\tilde{\alpha}}^{\prime} =0$, $d^\dagger {\tilde{\alpha}}^{\prime} = 0$, ${\tilde{d^\dagger}} {\tilde{\alpha}}^{\prime} =0$ and $\tau {\tilde{\alpha}}^{\prime} = 0$ -- again in complete agreement with the results in {\cite{keyMet}} for fields with $p=q$. For the $p>q$, $k=-1$ solution (with $\alpha =0$) one also has the physical gauge equation $\Delta_{[p,q]} \, A =0$ which again agrees with {\cite{keyMet}} in anti-de Sitter space (for $q>0$). For the $p>q$, $k=p-q$ solution (with ${\tilde{\alpha}} =0$) however, one has the physical gauge equation $\Delta^{\prime}_{[p,q]} \, A =0$ which does not agree with {\cite{keyMet}} in anti-de Sitter space unless $q=0$ (i.e. except for the case of $p$-form gauge fields). This is perhaps to be expected since the physical equations for a single $p>q>0$ gauge field found here depend on whether one began with the $k=-1$ or $k=p-q$ solution whereas the physical gauge equations for a single $p>q>0$ gauge field in {\cite{keyMet}} are unique and correspond to the $k=-1$ solution here. It is for a $q=0$ $p$-form gauge field that the $k=p-q$ physical gauge equations derived here correspond to those for a $p$-form gauge field in {\cite{keyMet}}. A further subtlety is that the physical gauge equations in {\cite{keyMet}} for $p>q>0$ gauge fields are invariant under any gauge transformations whose two independent parameters satisfy appropriate physical gauge equations themselves. In contrast, the $p>q>0$ solutions are derived here from an action which is only invariant under a restricted set of one-parameter gauge transformations. The way to reobtain the Metsaev type physical gauge equations for general $p>q>0$ gauge fields in anti-de Sitter space from a fully gauge-invariant action will be described in the next section.       


\section{Massive gauge fields on ${\cal{M}}_D ( \Omega )$}

We now consider adding extra fields to construct fully gauge-invariant actions for general gauge fields of type $[p,q]$. The inclusion of an arbitrary mass term for the original field $A \, \in \, X^{[p,q]}$ explicitly breaks gauge invariance and so the extra fields can be viewed as the appropriate compensator fields required to restore gauge invariance. This section is intended as a generalisation of the work {\cite{keyZin}} in order to describe gauge-invariant actions for gauge fields corresponding to any Young tableaux with two columns. In curved space, many of the features we find like $\Omega$-dependent mass bounds and the existence of only partially massless limits agree precisely with the examples given in {\cite{keyZin}} and are reminiscent of similar studies of massive totally symmetric tensor gauge theories in {\cite{keyDes1}}, {\cite{keyZin2}}, {\cite{keyDes2}}. The physical gauge equations for these theories are found in certain partially massless limits and complete agreement is found with the results of {\cite{keyMet}} in anti-de Sitter space. We also discuss the unitarity of the massive theory in flat and anti-de Sitter space.      


\subsection{Massive bi-form gauge theory in flat space}

We begin by describing the construction of gauge-invariant actions for general type $[p,q]$ massive gauge fields in flat space. Consider the action for a massive gauge field $A\, \in\, X^{[p,q]}$ of mass $m$ in flat space, given by
%
%
\begin{equation}
{\textsf{S}}_A - m^2 \, {\textsf{M}}_A
\label{eq:5.1}
\end{equation}
This is evidently not gauge-invariant under $\delta A = {\cal{Y}}_{[p,q]} \circ \left( d \alpha + {\tilde{d}} {\tilde{\alpha}} \right)$ for any $p \geq q$. Consider first the case of $p>q$. Since the mass term in ({\ref{eq:5.1}}) breaks all gauge symmetry then one must introduce the $GL(D,{\mathbb{R}})$-irreducible compensator fields $B\,\in\, X^{[p-1,q]}$ and ${\tilde{B}}\,\in\, X^{[p,q-1]}$ for the gauge parameters $\alpha \,\in\, X^{[p-1,q]}$ and ${\tilde{\alpha}} \,\in\, X^{[p,q-1]}$ respectively. For $p=q$ recall that one can write the gauge transformation in terms of only one parameter, ${\tilde{\alpha}}$ say, and therefore in that case one needs only one compensator field ${\tilde{B}}\,\in\, X^{[p,p-1]}$ (whilst effectively setting $B=0$). 

Both the compensator fields are gauge fields themselves, however, and so have their own gauge transformations     
%
%
\begin{eqnarray}
\delta B &=& {\cal{Y}}_{[p-1,q]} \circ \left( d \beta + {\tilde{d}} {\tilde{\beta}} \right) \nonumber \\
\delta {\tilde{B}} &=& {\cal{Y}}_{[p,q-1]} \circ \left( d \gamma + {\tilde{d}} {\tilde{\gamma}} \right)
\label{eq:5.2}
\end{eqnarray}
for any parameters $\beta \,\in\, X^{[p-2,q]}$, ${\tilde{\beta}} \,\in\, X^{[p-1,q-1]}$, $\gamma \,\in\, X^{[p-1,q-1]}$ and ${\tilde{\gamma}} \,\in\, X^{[p,q-2]}$. It may seem that one should then introduce four more gauge fields as compensator fields for each of these parameters. However, one must take into account that the original gauge transformation $\delta A$ vanishes identically for gauge parameters
%
%
\begin{eqnarray}
\alpha &=& {\cal{Y}}_{[p-1,q]} \circ \left( d \xi + {\tilde{d}} {\tilde{\xi}} \right) \nonumber \\
{\tilde{\alpha}} &=& {\cal{Y}}_{[p,q-1]} \circ \left( d \zeta + {\tilde{d}} {\tilde{\zeta}} \right)
\label{eq:5.3}
\end{eqnarray}
where $\xi \,\in\, X^{[p-2,q]}$, ${\tilde{\zeta}} \,\in\, X^{[p,q-2]}$ and ${\tilde{\xi}} = - \zeta \,\in\, X^{[p-1,q-1]}$. This means that these parts of the general gauge transformation for $A$ are not responsible for any symmetry breaking. Therefore one does not need to introduce compensator fields for $\beta$ and ${\tilde{\gamma}}$ in ({\ref{eq:5.2}}) and only one independent compensator field $C\,\in\, X^{[p-1,q-1]}$ for both ${\tilde{\beta}}$ and $\gamma$, with gauge transformation
%
%
\begin{equation}
\delta C \;\; =\;\; {\cal{Y}}_{[p-1,q-1]} \circ \left( d \varepsilon + {\tilde{d}} {\tilde{\varepsilon}} \right)
\label{eq:5.4}
\end{equation}
for any parameters $\varepsilon \,\in\, X^{[p-2,q-1]}$ and ${\tilde{\varepsilon}} \,\in\, X^{[p-1,q-2]}$. Repeating the procedure above, one finds that no more compensator fields are required for this gauge symmetry since $\delta A$ vanishes identically for gauge parameters $\alpha = d {\tilde{d}} \varsigma$, ${\tilde{\alpha}} = d {\tilde{d}} {\tilde{\varsigma}}$ in terms of any $\varsigma \,\in\, X^{[p-2,q-1]}$ and ${\tilde{\varsigma}} \,\in\, X^{[p-1,q-2]}$. The gauge transformations ({\ref{eq:5.4}}) are thus not responsible for any symmetry breaking. 
 
To summarise, for $p>q$ there are four fields in total $(A,B,{\tilde{B}},C)$ with eight independent gauge parameters $( \alpha , {\tilde{\alpha}} , \beta , {\tilde{\beta}} , \gamma , {\tilde{\gamma}} , \varepsilon , {\tilde{\varepsilon}} )$. For $p=q$ then there are three fields which can be taken to be $(A,{\tilde{B}},C)$ (by putting $B=0$ in the set above) and four independent gauge parameters $( {\tilde{\alpha}} , \gamma , {\tilde{\gamma}} , {\tilde{\varepsilon}} )$ (by putting $(  \alpha , \beta , {\tilde{\beta}} , \varepsilon )$ all equal to zero
\footnote{One must set $\beta$ and ${\tilde{\beta}}$ equal to zero for consistency with $B=0$ whilst $\alpha$ and $\varepsilon$ are set to zero since $p=q$ and so $p-1=q-1$.}
). 

For $p>q$, the appropriate gauge transformations for these fields, in terms of all the gauge parameters, are given by
%
%
\begin{eqnarray}
\delta A &=& {\cal{Y}}_{[p,q]} \circ \left( d \alpha + {\tilde{d}} {\tilde{\alpha}} \right) - {m \over (D-p-q)} \; \eta \left( \left( {p-q \over p-q+1} \right) {\tilde{\beta}} \pm {\sqrt{p-q+2 \over p-q+1}} \gamma \right) \nonumber \\
\nonumber \\
\delta B &=& {\cal{Y}}_{[p-1,q]} \circ \left( d \beta + {\tilde{d}} {\tilde{\beta}} \right) +m\, \alpha \pm {m \over {\sqrt{(D-p-q)(D-p-q+1)}}} \, \eta \varepsilon \label{eq:5.5} \\
\nonumber \\
\delta {\tilde{B}} &=& {\cal{Y}}_{[p,q-1]} \circ \left( d \gamma + {\tilde{d}} {\tilde{\gamma}} \right) +m\, {\sqrt{p-q+1 \over p-q+2}} \left( \pm {\tilde{\alpha}} + {1 \over {\sqrt{(D-p-q)(D-p-q+1)}}} \, \eta {\tilde{\varepsilon}} \right) \nonumber \\
\nonumber \\
\delta C &=& {\cal{Y}}_{[p-1,q-1]} \circ \left( d \varepsilon + {\tilde{d}} {\tilde{\varepsilon}} \right) - m\, {\sqrt{D-p-q+1 \over D-p-q}} \left( \pm \left( {p-q \over p-q+1} \right) {\tilde{\beta}} + {\sqrt{p-q+2 \over p-q+1}} \gamma \right) 
\nonumber
\end{eqnarray}
with coefficients chosen precisely so that the action
%
%
\begin{eqnarray}
&&{\textsf{S}}_A + {\textsf{S}}_B + {\textsf{S}}_{\tilde{B}} + {\textsf{S}}_C - m^2 \, {\textsf{M}}_A + m^2  \left( {D-p-q+2 \over D-p-q} \right) {\textsf{M}}_C \nonumber \\
\nonumber \\
&&+2m \, {\left( A , \sum_{n=0}^{q} {{(-1)}^n \over {(n!)}^2} \, \eta^n \tau^n \left[ dB \pm {\sqrt{p-q+2 \over p-q+1}} {\tilde{d}} {\tilde{B}} \right] \right)}_{p,q} \label{eq:5.6} \\ 
\nonumber \\
&&+2m \, {\sqrt{D-p-q+1 \over D-p-q}} {\left( C , \sum_{n=0}^{q-1} {{(-1)}^n \over {(n+1) (n!)}^2} \, \eta^n \tau^{n+1} \left[ \pm dB + {\sqrt{p-q+2 \over p-q+1}} {\tilde{d}} {\tilde{B}} \mp m\, A \right] \right)}_{p-1,q-1} 
\nonumber 
\end{eqnarray}
is gauge-invariant. Notice that these formulas are valid for any value of $m\, \in\, {\mathbb{R}}$ so that one can take the massless limit $m \rightarrow 0$ of ({\ref{eq:5.6}}) to obtain the decoupled actions for the four free fields $(A,B,{\tilde{B}},C)$ in flat space. The fact that ({\ref{eq:5.6}}) is gauge-invariant for either of the various plus or minus sign choices can just be taken as a fact for the moment -- its origin will become clearer when we discuss massive fields in curved space. One can obtain the corresponding gauge transformations and invariant action for the $p=q$ case by simply putting $(B , \alpha , \beta , {\tilde{\beta}} ,\varepsilon)$ all equal to zero in ({\ref{eq:5.5}}) and ({\ref{eq:5.6}}). Recall that the free field equation for a single field $A \, \in\, X^{[p,q]}$ is non-trivial in dimension $D \geq p+q+2$. If this bound is therefore assumed in ({\ref{eq:5.5}}) and ({\ref{eq:5.6}}) then, since $p \geq q$, none of the terms ever becomes imaginary or singular. The results above agree with those derived in {\cite{keyZin}} for tensor gauge fields of type $[2,1]$, $[3,1]$ and $[2,2]$. 

The process of going to physical gauge for this system of fields $(A,B,{\tilde{B}},C)$ involves \lq gauging away' the three compensator fields $(B,{\tilde{B}},C)$, leaving a set of equations in terms of the fundamental field $A$ only. More precisely, given the four gauge-invariant field equations for $(A,B,{\tilde{B}},C)$ derived from ({\ref{eq:5.6}}) then one can redefine the fields by a particular (field-dependent) gauge transformation (of the form ({\ref{eq:5.5}})) such that the four equations of motion can be expressed in terms of the (redefined) fundamental field $A$ only. 
\footnote{This mechanism is best demonstrated explicitly via a simple example. Consider the equation of motion $\partial^{\mu} F_{\mu\nu} - m^2 A_{\nu} =0$ for a massive Maxwell field $A_{\mu}$ (with field strength $F_{\mu\nu} = \partial_{\mu} A_{\nu} - \partial_{\nu} A_{\mu}$ and mass $m$). Taking the divergence of this equation, one finds that it implies $\partial^{\mu} A_{\mu} =0$ for $m \neq 0$. (This is different from the massless theory where $\partial^{\mu} A_{\mu} =0$ does not follow from Maxwell's equations but from a choice of gauge.) Putting $\partial^{\mu} A_{\mu} =0$ back into the massive field equation then implies $( \square - m^2 ) A_{\mu} =0$ which are the correct Maxwell-Proca equations for a massive spin-1 field. Evidently this massive theory is not gauge-invariant but can be made so by introducing a compensating scalar field $\phi$. The massive Lagrangian $-{1 \over 4} F_{\mu\nu} F^{\mu\nu} - {1 \over 2} m^2 A_{\mu} A^{\mu} - {1 \over 2} \partial_{\mu} \phi \partial^{\mu} \phi - m \phi \partial^{\mu} A_{\mu}$ is then invariant under the gauge transformations $\delta A_{\mu} = \partial_{\mu} \lambda$, $\delta \phi = m \lambda$. The equations of motion for $( A_{\mu} , \phi )$ are respectively $\partial^{\mu} F_{\mu\nu} - m^2 A_{\nu} + m \partial_{\nu} \phi =0$ and $\square\, \phi -m \partial^{\mu} A_{\mu} =0$, which become the correct Maxwell-Proca equations $( \square - m^2 ) A_{\mu} =0$ and $\partial^{\mu} A_{\mu} =0$ on redefining $A_{\mu} \rightarrow A_{\mu} + {1 \over m} \partial_{\mu} \phi$. This is simply a field-dependent gauge transformation of $A_{\mu}$ with parameter ${1 \over m} \phi$. A similar (but more complicated) story follows for the case we consider above. For example, one can gauge away $B$ and ${\tilde{B}}$ by redefining $A \rightarrow A + \delta A$ for field-dependent gauge parameters $\alpha = {1 \over m} B$ and ${\tilde{\alpha}} = \pm {1 \over m} {\sqrt{p-q+2 \over p-q+1}} {\tilde{B}}$.}
In terms of this redefined field $A$, the four equations of motion become the physical gauge equations
%
%
\begin{equation}
( \square - m^2 )\, A \;\; =\;\; 0 \quad , \quad d^\dagger A \;\; =\;\; 0 \quad , \quad {\tilde{d^\dagger}} A \;\; =\;\; 0 \quad , \quad \tau A \;\; =\;\; 0 
\label{eq:5.6b}
\end{equation}
where each of the four equations above is simply the redefined form of the field equation for each of the original fields $(A,B,{\tilde{B}},C)$ respectively.


\subsection{Massive bi-form gauge theory on ${\cal{M}}_D ( \Omega )$}

We now repeat the procedure above for a massive gauge field $A\,\in\, X^{[p,q]}$ on ${\cal{M}}_D ( \Omega )$. The analysis of compensator fields in flat space indicates that we should again consider the set of fields $(A,B,{\tilde{B}},C)$ on ${\cal{M}}_D ( \Omega )$ with gauge parameters $( \alpha , {\tilde{\alpha}} , \beta , {\tilde{\beta}} , \gamma , {\tilde{\gamma}} , \varepsilon , {\tilde{\varepsilon}} )$. The results will again be valid for any $p \geq q >0$, with the $p=q$ case obtained by setting the field $B$ and the gauge parameters $( \alpha , \beta , {\tilde{\beta}} ,\varepsilon)$ equal to zero.

The appropriate (covariant) gauge transformations for these fields on ${\cal{M}}_D ( \Omega )$, in terms of all the gauge parameters, are given by
%
%
\begin{eqnarray}
\delta A &=& {\cal{Y}}_{[p,q]} \circ \left( d \alpha + {\tilde{d}} {\tilde{\alpha}} \right) - {1 \over (D-p-q)} \; g \left( \mu_1 \left( {p-q \over p-q+1} \right) {\tilde{\beta}} + \mu_2 {\sqrt{p-q+2 \over p-q+1}} \gamma \right) \nonumber \\
\nonumber \\
\delta B &=& {\cal{Y}}_{[p-1,q]} \circ \left( d \beta + {\tilde{d}} {\tilde{\beta}} \right) + \mu_1 \, \alpha + {\mu_2 \over {\sqrt{(D-p-q)(D-p-q+1)}}} \, g \varepsilon \label{eq:5.7} \\
\nonumber \\
\delta {\tilde{B}} &=& {\cal{Y}}_{[p,q-1]} \circ \left( d \gamma + {\tilde{d}} {\tilde{\gamma}} \right) + {\sqrt{p-q+1 \over p-q+2}} \left( \mu_2 {\tilde{\alpha}} + {\mu_1 \over {\sqrt{(D-p-q)(D-p-q+1)}}} \, g {\tilde{\varepsilon}} \right) \nonumber \\
\nonumber \\
\delta C &=& {\cal{Y}}_{[p-1,q-1]} \circ \left( d \varepsilon + {\tilde{d}} {\tilde{\varepsilon}} \right) - {\sqrt{D-p-q+1 \over D-p-q}} \left( \mu_2 \left( {p-q \over p-q+1} \right) {\tilde{\beta}} + \mu_1 {\sqrt{p-q+2 \over p-q+1}} \gamma \right) 
\nonumber
\end{eqnarray}
with coefficients chosen so that the action
%
%
\begin{eqnarray}
&&{\textsf{S}}_A + {\textsf{S}}_B + {\textsf{S}}_{\tilde{B}} + {\textsf{S}}_C - ( {\mu_2}^2 - \Omega\, (D-p-q) ) \, {\textsf{M}}_A - \Omega\, (p-q-1)(D-p-q+1) \, {\textsf{M}}_B \nonumber \\
\nonumber \\
&&+ \Omega\, (D-p-q+1) \, {\textsf{M}}_{\tilde{B}} + ( {\mu_1}^2 + \Omega\, (D-p-q) )  \left( {D-p-q+2 \over D-p-q} \right) {\textsf{M}}_C \nonumber \\
\nonumber \\
&&+2 \, {\left( A , \sum_{n=0}^{q} {{(-1)}^n \over {(n!)}^2} \, g^n \tau^n \left[ \mu_1 dB + \mu_2 {\sqrt{p-q+2 \over p-q+1}} {\tilde{d}} {\tilde{B}} \right] \right)}_{p,q} \label{eq:5.8} \\ 
\nonumber \\
&&+2 \, {\sqrt{D-p-q+1 \over D-p-q}} {\left( C , \sum_{n=0}^{q-1} {{(-1)}^n \over {(n+1) (n!)}^2} \, g^n \tau^{n+1} \left[ \mu_2 dB + \mu_1 {\sqrt{p-q+2 \over p-q+1}} {\tilde{d}} {\tilde{B}} - \mu_1 \mu_2 \, A \right] \right)}_{p-1,q-1} \nonumber \\ 
\nonumber 
\end{eqnarray}
is gauge-invariant on ${\cal{M}}_D ( \Omega )$. Gauge invariance of ({\ref{eq:5.8}}) also demands that the two real numbers $\mu_1$ and $\mu_2$ be related such that
%
%
\begin{equation}
{\mu_1}^2 \;\; =\;\; {\mu_2}^2 - \Omega\, (p-q+1)(D-p-q)
\label{eq:5.9}
\end{equation}
Consequently there is only one unfixed \lq mass' parameter in the equations above. For general gauge parameters in the transformations ({\ref{eq:5.7}}) (i.e. which satisfy no special properties other than $GL(D,{\mathbb{R}})$-irreducibility), the gauge-invariant action ({\ref{eq:5.8}}) is unique, with all coefficients in ({\ref{eq:5.7}}) and ({\ref{eq:5.8}}) being fixed by gauge-invariance. 

It is clear that in the flat space $\Omega \rightarrow 0$ limit the expressions above reduce to those described in the previous subsection with $m^2 = {\mu_1}^2 = {\mu_2}^2$ (so that $\mu_1 = \pm \mu_2 =m$). In curved space however there are many qualitatively different features of the model that we will now describe. Note that the cosmological constant has the effect of shifting the values of all the mass-like terms in ({\ref{eq:5.8}}) so that, in particular, $B$ and ${\tilde{B}}$ now have non-vanishing mass-like terms.

The most striking feature of ({\ref{eq:5.8}}), for $\Omega \neq 0$, is the absence of a limit in which {\textit{both}} $\mu_1$ and $\mu_2$ tend to zero. This is, of course, due to ({\ref{eq:5.9}}) which constrains these parameters to lie on hyperbolic curves in the  $\mu_1$--$\mu_2$ plane for the $\Omega \neq 0$ geometries, which asymptotically approach the straight lines $\mu_1 = \pm \mu_2$ at infinity (these two lines correspond to the allowed values of $\mu_1$ and $\mu_2$ in Minkowski space) [see Figure 1]
\vspace*{.2in}

%
%
\setlength{\unitlength}{.5in}
\begin{picture}(8,8)
\hspace{1.2in}\includegraphics[width=4in,height=4in]{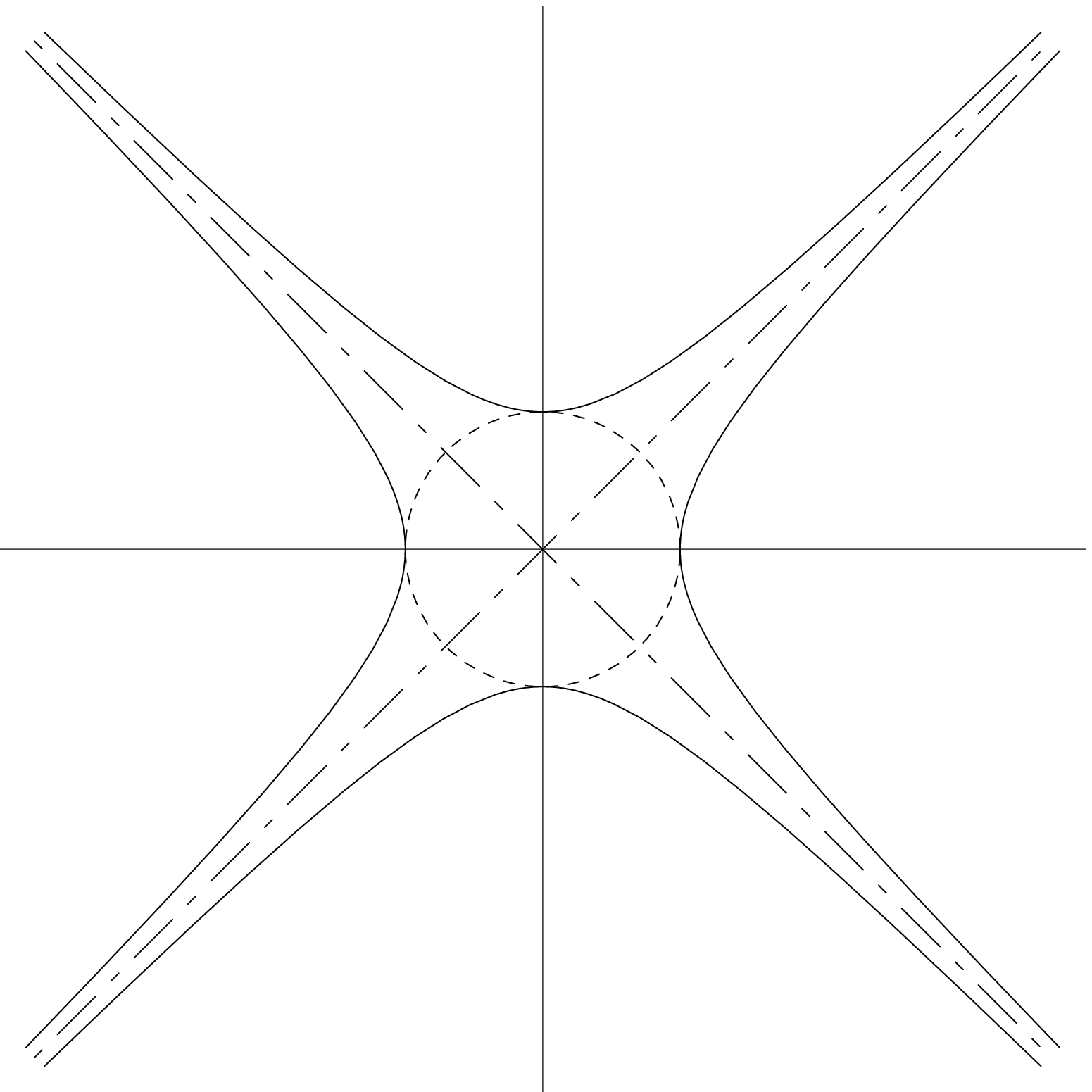}
\hspace*{-3.85in} 
\put(3.45,8.15){$\mu_1$}
\put(2.75,6.2){\footnotesize $\Omega < 0\;$ anti-de Sitter}
\put(0.3,4.1){\footnotesize $\Omega >0\;$ de Sitter}
\put(5,4.1){\footnotesize $\Omega >0\;$ de Sitter}
\put(7.7,3.95){$\mu_2$}
\put(2.75,1.65){\footnotesize $\Omega < 0\;$ anti-de Sitter}
\end{picture}
\begin{center}
{\footnotesize Figure 1 : Curves representing the allowed values of $\mu_1$ and $\mu_2$ for the ${adS}_D$, ${dS}_D$ and Minkowski geometries. The two diagonal lines $\mu_1 = \pm \mu_2$ correspond to Minkowski space where $\Omega =0$. Given a fixed absolute value for the cosmological constant then the dashed circle has radius $\sqrt{|\Omega|(p-q+1)(D-p-q)}$.} 
\end{center}
The two diagonal lines $\mu_1 = \pm \mu_2$ partition the $\mu_1$--$\mu_2$ plane into four sectors. For the ${adS}_D$ geometry, the hyperbola lies in the upper and lower sectors only whilst, for the ${dS}_D$ geometry, it lies only in the left and right sectors. Given a fixed absolute value of the cosmological constant then for either of the $\Omega \neq  0$ geometries, neither of the parameters $\mu_1$ and $\mu_2$ can take values within a circle of radius ${\sqrt{| \Omega | \, (p-q+1)(D-p-q)}}$ in the $\mu_1$--$\mu_2$ plane. For ${adS}_D$, one can take $\mu_2 =0$ (at $\mu_1 = \pm {\sqrt{-\Omega\, (p-q+1)(D-p-q)}}$) but $\mu_1 \neq 0$ everywhere. For ${dS}_D$, one can take $\mu_1 =0$ (at $\mu_2 = \pm {\sqrt{\Omega\, (p-q+1)(D-p-q)}}$) but $\mu_2 \neq 0$ everywhere. One can, of course, take both $\mu_1 = \pm \mu_2 =m=0$ in Minkowski space.

Taking the $\mu_2 \rightarrow 0$ limit in ${adS}_D$ decouples the system with fields $(A,B,{\tilde{B}},C)$ into two subsystems with fields $( A,B )$ and $( {\tilde{B}} ,C)$ (for $p=q$ then one has the single (massless) field $A$ decoupled from $( {\tilde{B}} ,C)$). The decoupled fields have the gauge transformations
%
%
\begin{eqnarray}
\delta A &=& {\cal{Y}}_{[p,q]} \circ \left( d \alpha + {\tilde{d}} {\tilde{\alpha}} \right) \mp {\sqrt{-\Omega \, {(p-q)}^2 \over (p-q+1)(D-p-q)}} \, g {\tilde{\beta}} \nonumber \\
\nonumber \\
\delta B &=& {\cal{Y}}_{[p-1,q]} \circ \left( d \beta + {\tilde{d}} {\tilde{\beta}} \right) \pm {\sqrt{-\Omega\, (p-q+1)(D-p-q)}} \, \alpha  \label{eq:5.10} \\
\nonumber \\
\delta {\tilde{B}} &=& {\cal{Y}}_{[p,q-1]} \circ \left( d \gamma + {\tilde{d}} {\tilde{\gamma}} \right) \pm {\sqrt{ -\Omega \, {(p-q+1)}^2  \over (p-q+2)(D-p-q+1)}} \, g {\tilde{\varepsilon}} \nonumber \\
\nonumber \\
\delta C &=& {\cal{Y}}_{[p-1,q-1]} \circ \left( d \varepsilon + {\tilde{d}} {\tilde{\varepsilon}} \right) \mp {\sqrt{-\Omega\, (p-q+2)(D-p-q+1)}} \, \gamma 
\nonumber
\end{eqnarray}
under which the decoupled actions
%
%
\begin{eqnarray}
&&{\textsf{S}}_A + {\textsf{S}}_B + \Omega\, (D-p-q) \, {\textsf{M}}_A - \Omega\, (p-q-1)(D-p-q+1) \, {\textsf{M}}_B 
\nonumber \\
\nonumber \\
&&\pm 2 \, {\sqrt{-\Omega \, (p-q+1)(D-p-q)}} {\left( A , \sum_{n=0}^{q} {{(-1)}^n \over {(n!)}^2} \, g^n \tau^n \, dB \right)}_{p,q} \label{eq:5.11} \\ 
\nonumber 
\end{eqnarray}
and
%
%
\begin{eqnarray}
&&{\textsf{S}}_{\tilde{B}} + {\textsf{S}}_C + \Omega\, (D-p-q+1) \, {\textsf{M}}_{\tilde{B}} - \Omega\, (p-q)(D-p-q+2) \, {\textsf{M}}_C \nonumber \\
\nonumber \\
&&\pm 2 \, {\sqrt{-\Omega\, (p-q+2)(D-p-q+1)}} {\left( C , \sum_{n=0}^{q-1} {{(-1)}^n \over {(n+1) (n!)}^2} \, g^n \tau^{n+1} {\tilde{d}} {\tilde{B}} \right)}_{p-1,q-1} \label{eq:5.12} \\ 
\nonumber
\end{eqnarray}
are individually invariant on ${adS}_D$. Notice that the action and gauge transformations for the pair of fields $(A,B)$ are the same as those for $({\tilde{B}},C)$ upon replacing $q$ by $q-1$. Each pair of fields could therefore be described as massless, in the sense that the action for each pair is both invariant under all the gauge symmetries of the massless action for the corresponding pair of fields in flat space and reduces to the sum of massless actions for each field in the flat space limit. This description is, of course, only possible because of the coupling term between the two elements of each pair of fields. As found in section 3, if no such couplings are present then only a partially massless action can be constructed for a general gauge field with $p>q>0$. An example of the structure above is found in {\cite{keyMetVas}} where they discover that the action for a single $[2,1]$ \lq hook' gauge field (what we call $A \in X^{[2,1]}$) must be supplemented with that for an auxiliary $[1,1]$ graviton-like gauge field (what we call $B \in X^{[1,1]}$) (plus coupling terms between them) in anti-de Sitter space in order to preserve all the gauge symmetries of the corresponding massless theory in flat space. Moreover, note that in the $\mu_2 \rightarrow 0$ limit above, the decoupled pair $( A,B )$ consists of two two-column Young tableaux (since $p > q>0$) whilst the pair $( {\tilde{B}} ,C)$ consist of either two two-column (when $q>1$) or two one-column (when $q=1$) Young tableaux. A (one-column,two-column) mixed pair never occurs. This apartheid between form fields and more general Young tableaux is discussed in detail in {\cite{keyMetVas}} for massless gauge theories in anti-de Sitter space.      

The equations of motion derived from ({\ref{eq:5.11}}) for the decoupled pair of fields $(A,B)$ in physical gauge are
\footnote{The procedure for going to physical gauge here is as follows. One can use the $\alpha \, \in \, X^{[p-1,q]}$ gauge transformations in ({\ref{eq:5.10}}) to gauge away the compensator field $B \, \in \, X^{[p-1,q]}$ in the equations of motion. In terms of the redefined field $A$, the field equations for $(A,B)$ become $\Delta_{[p,q]} \, A  =0$ and $d^\dagger A =0$ respectively. Residual symmetries corresponding to transformations of the original fields with gauge parameters ${\tilde{\alpha}} \, \in \, X^{[p,q-1]}$ and ${\tilde{\beta}} \, \in \, X^{[p-1,q-1]}$ can then be used to set ${\tilde{d^\dagger}} A =0$ and $\tau A =0$ respectively (on-shell). The remaining $\beta \, \in \, X^{[p-2,q]}$ gauge parameter generates a trivial symmetry of the redefined theory above.}
%
%
%
\begin{equation}
\Delta_{[p,q]} \, A \;\; =\;\; 0 \quad , \quad d^\dagger A \;\; =\;\; 0 \quad , \quad {\tilde{d^\dagger}} A \;\; =\;\; 0 \quad , \quad \tau A \;\; =\;\; 0  
\label{eq:5.12a} 
\end{equation}
(those for $({\tilde{B}},C)$ follow from this analysis by replacing $q$ by $q-1$). The physical gauge equations for $A \, \in \, X^{[p,q]}$ in this partially massless limit in ${adS}_D$ agree with those given in {\cite{keyMet}} for general gauge fields with $p \geq q >0$ and are invariant under gauge transformations with parameters $\alpha$ and ${\tilde{\alpha}}$ satisfying $\Delta^{\prime}_{[p-1,q]} \, \alpha =0$ and $\Delta_{[p,q-1]} \, {\tilde{\alpha}} =0$ (in addition to their also being annihilated by $d^\dagger$, ${\tilde{d^\dagger}}$ and $\tau$). 

To consider the unitarity of the theory on ${adS}_D$, we now reintroduce a non-vanishing mass parameter $\mu_2$. In physical gauge, the equations of motion derived from ({\ref{eq:5.8}}) become
%
%
%
\begin{equation}
( \Delta_{[p,q]} - {\mu_2}^2 ) \, A \;\; =\;\; 0 \quad , \quad d^\dagger A \;\; =\;\; 0 \quad , \quad {\tilde{d^\dagger}} A \;\; =\;\; 0 \quad , \quad \tau A \;\; =\;\; 0  
\label{eq:5.12b} 
\end{equation}
The Laplacian equation in ({\ref{eq:5.12b}}) can be solved using standard techniques in anti-de Sitter space (for example, see {\cite{keyholog}} and references therein). The energy of a given solution is determined by the two roots of a single quadratic equation (satisfied by the eigenvalues of the Laplace equation at the boundary of ${adS}_D$). In particular, the energy of a solution is only real provided the two roots are also real. Real energy therefore requires the argument of the square root in the expression for the two roots to be non-negative -- this is the Breitenlohner-Freedman unitarity bound {\cite{keybrefre}}. Solving the Laplacian equation in ({\ref{eq:5.12b}}) leads to the Breitenlohner-Freedman bound
%
%
\begin{equation}
-{\Omega \over 4} {(D+1-2q)}^2 + {\mu_2}^2 \;\; \geq \;\; 0 
\label{eq:5.12c} 
\end{equation}
for the fundamental field $A$.
\footnote{This follows from the Breitenlohner-Freedman bound $-{\Omega \over 4} {(D+1-2q)}^2 + {({m_{[p,q]}})}^2 \geq 0$  derived from the equation $( \Delta_{[p,q]} - {({m_{[p,q]}})}^2 ) A = 0$ for a tensor field $A \,\in\, X^{[p,q]}$ on ${adS}_D$. The bound $-{\Omega \over 4} {(D-1-2p)}^2 + {({m^{\prime}_{[p,q]}})}^2 \geq 0$ follows from the equation $( \Delta^{\prime}_{[p,q]} - {({m^{\prime}_{[p,q]}})}^2 ) A = 0$. As noted in section 3, it is the primed equation that is satisfied by physical $p$-form gauge fields (with $q=0$). The general mass bound on ${({m^{\prime}_{[p,q]}})}^2$ is independent of $q$ and is indeed identical to that for a massive $p$-form gauge field on ${adS}_D$. For example, if $p=q=0$, we have ${({m^{\prime}_{[0,0]}})}^2 \geq {\Omega \over 4} {(D-1)}^2$ which is the well known mass bound for scalar fields on ${adS}_D$.}
Since we have assumed all masses to be real throughout this paper then ({\ref{eq:5.12c}}) is satisfied identically on ${adS}_D$. As is typical in anti-de Sitter space however, if one analytically continues this assumption so that negative values of mass-squared are allowed then it is clear that the Breitenlohner-Freedman bound above can be saturated.
        
Taking the $\mu_1 \rightarrow 0$ limit in ${dS}_D$ again decouples the system with fields $(A,B,{\tilde{B}},C)$ into two subsystems but with fields $( A, {\tilde{B}} )$ and $( B ,C)$ (for $p=q$ then one has the single (massless) field $C$ decoupled from $( A, {\tilde{B}} )$). The decoupled fields have the gauge transformations
%
%
\begin{eqnarray}
\delta A &=& {\cal{Y}}_{[p,q]} \circ \left( d \alpha + {\tilde{d}} {\tilde{\alpha}} \right) \mp {\sqrt{\Omega\, (p-q+2) \over D-p-q}} \, g\gamma  \nonumber \\
\nonumber \\
\delta B &=& {\cal{Y}}_{[p-1,q]} \circ \left( d \beta + {\tilde{d}} {\tilde{\beta}} \right) \pm {\sqrt{\Omega\, (p-q+1) \over D-p-q+1 }} \, g \varepsilon \label{eq:5.13} \\
\nonumber \\
\delta {\tilde{B}} &=& {\cal{Y}}_{[p,q-1]} \circ \left( d \gamma + {\tilde{d}} {\tilde{\gamma}} \right) \pm {\sqrt{\Omega\, {(p-q+1)}^2 (D-p-q) \over p-q+2}} \, {\tilde{\alpha}}  \nonumber \\
\nonumber \\
\delta C &=& {\cal{Y}}_{[p-1,q-1]} \circ \left( d \varepsilon + {\tilde{d}} {\tilde{\varepsilon}} \right) \mp {\sqrt{\Omega\, {(p-q)}^2 (D-p-q+1) \over p-q+1}} \, {\tilde{\beta}}  
\nonumber
\end{eqnarray}
under which the decoupled actions
%
%
\begin{eqnarray}
&&{\textsf{S}}_A + {\textsf{S}}_{\tilde{B}} - \Omega\, (p-q)(D-p-q) \, {\textsf{M}}_A + \Omega\, (D-p-q+1) \, {\textsf{M}}_{\tilde{B}} 
\nonumber \\
\nonumber \\
&&\pm 2 \, {\sqrt{\Omega \, (p-q+2)(D-p-q)}} {\left( A , \sum_{n=0}^{q} {{(-1)}^n \over {(n!)}^2} \, g^n \tau^n \, {\tilde{d}} {\tilde{B}} \right)}_{p,q} \label{eq:5.14} \\ 
\nonumber 
\end{eqnarray}
and
%
%
\begin{eqnarray}
&&{\textsf{S}}_B + {\textsf{S}}_C - \Omega\, (p-q-1)(D-p-q+1) \, {\textsf{M}}_B + \Omega\, (D-p-q+2) \, {\textsf{M}}_C \nonumber \\
\nonumber \\
&&\pm 2 \, {\sqrt{\Omega\, (p-q+1)(D-p-q+1)}} {\left( C , \sum_{n=0}^{q-1} {{(-1)}^n \over {(n+1) (n!)}^2} \, g^n \tau^{n+1} dB \right)}_{p-1,q-1} \label{eq:5.15} \\ 
\nonumber
\end{eqnarray}
are individually invariant on ${dS}_D$. Notice that the action and gauge transformations for the pair of fields $(A,{\tilde{B}})$ are the same as those for $(B,C)$ upon replacing $p$ by $p-1$. Each pair of fields could again be described as massless. 

The equations of motion derived from ({\ref{eq:5.14}}) for the decoupled pair of fields $(A,{\tilde{B}})$ in physical gauge are
\footnote{The procedure for going to physical gauge here is as follows. One can use the ${\tilde{\alpha}} \, \in \, X^{[p,q-1]}$ gauge transformations in ({\ref{eq:5.13}}) to gauge away the compensator field ${\tilde{B}} \, \in \, X^{[p,q-1]}$ in the equations of motion. In terms of the redefined field $A$, the field equations for $(A,{\tilde{B}})$ become $\Delta^{\prime}_{[p,q]} \, A  =0$ and ${\tilde{d^\dagger}} A =0$ respectively. Residual symmetries corresponding to transformations of the original fields with gauge parameters $\alpha \, \in \, X^{[p-1,q]}$ and $\gamma \, \in \, X^{[p-1,q-1]}$ can then be used to set $d^\dagger A =0$ and $\tau A =0$ respectively (on-shell). The remaining ${\tilde{\gamma}} \, \in \, X^{[p,q-2]}$ gauge parameter generates a trivial symmetry of the redefined theory above.}
%
%
%
\begin{equation}
\Delta^{\prime}_{[p,q]} \, A \;\; =\;\; 0 \quad , \quad d^\dagger A \;\; =\;\; 0 \quad , \quad {\tilde{d^\dagger}} A \;\; =\;\; 0 \quad , \quad \tau A \;\; =\;\; 0  
\label{eq:5.15a} 
\end{equation}
(those for $(B,C)$ follow from this analysis by replacing $p$ by $p-1$). The physical gauge equations for $A$ in this partially massless limit in ${dS}_D$ are invariant under gauge transformations with parameters $\alpha$ and ${\tilde{\alpha}}$ satisfying $\Delta^{\prime}_{[p-1,q]} \, \alpha =0$ and $\Delta_{[p,q-1]} \, {\tilde{\alpha}} =0$ (in addition to their being annihilated by $d^\dagger$, ${\tilde{d^\dagger}}$ and $\tau$).

Finally, it is worth noting that another gauge-invariant system exists on ${\cal{M}}_D ( \Omega )$, even though it cannot be obtained as a consistent limit of the action ({\ref{eq:5.8}}). To arrive at this system, begin by taking $p=q$ and then considering the transformations ({\ref{eq:5.7}}) and action ({\ref{eq:5.8}}) {\textit{without}} the constraint ({\ref{eq:5.9}}) on the two parameters $\mu_1$ and $\mu_2$. Evidently ({\ref{eq:5.8}}) is no longer gauge-invariant, however, by then taking $\mu_1 = \mu_2 =0$ with $B= {\tilde{B}} =0$ one obtains 
%
%
\begin{equation}
{(A, E_A )}_{p,p} + {(C, E_C )}_{p-1,p-1} 
\label{eq:5.16}
\end{equation}
which is simply the massless action for the $[p,p]$ and $[p-1,p-1]$ fields $A$ and $C$ on ${\cal{M}}_D ( \Omega )$, defined in ({\ref{eq:3.15c}}), which is invariant under the minimal gauge transformations $\delta A = {\cal{Y}}_{[p,p]} \circ \left( d \alpha + {\tilde{d}} {\tilde{\alpha}} \right)$ and $\delta C = {\cal{Y}}_{[p-1,p-1]} \circ \left( d \varepsilon + {\tilde{d}} {\tilde{\varepsilon}} \right)$.


\section{Arbitrary massive gauge fields on spaces of constant curvature}

Having described gauge fields corresponding to any Young tableaux with two columns on ${\cal{M}}_D ( \Omega )$, we conclude by making some remarks on how the structure is generalised to describe arbitrary representations of $GL(D,{\mathbb{R}})$. In particular we give a procedure for determining the appropriate set of fields necessary to make the massive system gauge-invariant. This description uses the generalised {\textit{multi-form}} construction. We will therefore very briefly review the parts of this construction that are relevant to the forthcoming discussion (a more detailed description (in flat space) can be found in {\cite{keythesis}}, {\cite{keydeMHul1}}, {\cite{keydeMHul2}}, {\cite{keyDVHen}}, {\cite{keyBekBou}}).  
   

\subsection{Multi-forms}

A {\textit{multi-form}} of order $N$ is a tensor field $T$ over ${\cal{M}}_D ( \Omega )$ that is an element of the $GL(D,{\mathbb{R}})$-reducible $N$-fold tensor product space of $p_i$-forms (where $i=1,...,N$), written 
%
%
\be
X^{p_1 ,..., p_N} \;\; := \;\; \Lambda^{p_1} \otimes ... \otimes \Lambda^{p_N}
\label{eq:4.1}
\ee
The components of $T$ are written $T_{{\mu}^1_1 ...{\mu}^1_{p_1} ...{\mu}^i_1 ...{\mu}^i_{p_i} ... {\mu}^N_{p_N} }$ and are taken to be totally antisymmetric in each set of $\{ {\mu}^i \}$ indices, such that
%
%
\be
T_{[ {\mu}^1_1 ...{\mu}^1_{p_1} ] ...[ {\mu}^i_1 ...{\mu}^i_{p_i} ]... [ {\mu}^N_{1} ... {\mu}^N_{p_N} ]} \;\; = \;\; T_{{\mu}^1_1 ...{\mu}^1_{p_1} ...{\mu}^i_1 ...{\mu}^i_{p_i} ... {\mu}^N_{1} ... {\mu}^N_{p_N} }
\label{eq:4.2}
\ee

The generalisation of the operations defined for bi-forms to multi-forms of order $N$ is then straightforward. There are $N$ inequivalent exterior derivatives 
%
%
\be
d^{(i)} \; :\; X^{p_1 ,...,p_i ,...p_N } \rightarrow X^{p_1 ,...,p_i +1 ,...,p_N }
\label{eq:4.4}
\ee
which are individually defined, by analogy with ({\ref{eq:2.5}}), as the exterior derivatives acting on the
$\Lambda^{p_i}$ form subspaces. 

There are $N$ inequivalent Hodge dual operations
%
%
\be
*^{(i)} \; :\; X^{p_1 ,...,p_i ,...p_N } \to X^{p_1 ,...,D- p_i ,...p_N }
\label{eq:4.7}
\ee
which, following ({\ref{eq:2.14}}), are defined to act as the Hodge duals on the individual $\Lambda^{p_i}$ form subspaces. This implies that ${*^{(i)}}^2 = (-1)^{1+ p_i (D-p_i )}$ (with no sum over $i$) with any two $*^{(i)}$  commuting. There exist $N(N-1)/2$ inequivalent trace operations
%
%
\be
\tau^{(ij)} \; :\; X^{p_1 ,...,p_i ,..., p_j ,...p_N } \rightarrow X^{p_1 ,...,p_i -1,..., p_j -1,...,p_N }
\label{eq:4.10}
\ee
defined, by analogy with ({\ref{eq:2.19}}), as the single trace between the $\Lambda^{p_i}$ and $\Lambda^{p_j}$ form subspaces using the inverse metric $g^{{\mu}^i_1 {\mu}^j_1}$. This allows one to define the \lq adjoint trace' operation
%
%
\be
\sigma^{(ij)} \; :=\; {(-1)}^{1+D( p_i +1)} *^{(i)} \tau^{(ij)} *^{(i)} \; :\; X^{p_1 ,...,p_i ,..., p_j ,...p_N } \rightarrow X^{p_1 ,...,p_i +1,..., p_j -1,...,p_N }
\label{eq:4.11}
\ee
associated with a given $\tau^{(ij)}$ (with no sum over $i$ or $j$).

Finally, there are $N(N-1)/2$ distinct operations
%
%
\be
g^{(ij)} : X^{p_1 ,...,p_i ,..., p_j ,...p_N} \rightarrow X^{p_1 ,...,p_i +1,..., p_j +1,...,p_N}
\label{eq:4.14}
\ee
defined such that for any $T \in X^{p_1 ,..., p_N}$ then $g^{(ij)} T$ has components \\
${( g^{(ij)} T)}_{{\mu}^1_1 ...{\mu}^1_{p_1} ...{\mu}^i_1 ...{\mu}^i_{p_i +1} ...{\mu}^j_1 ...{\mu}^j_{p_j +1} ... {\mu}^N_{p_N} } = ( p_i +1)( p_j +1) \, g_{\mu^i_1 \mu^j_1} T_{{\mu}^1_1 ...{\mu}^1_{p_1} ...{\mu}^i_2 ...{\mu}^i_{p_i +1} ... {\mu}^j_2 ...{\mu}^j_{p_j +1} ... {\mu}^N_{p_N} }$ (with implicit antisymmetrisation of each set of $\mu^i$ and $\mu^j$ indices).

Each of the exterior derivatives ({\ref{eq:4.4}}) is not nilpotent on ${\cal{M}}_D ( \Omega )$ and they do not commute with each other since
%
%
\begin{eqnarray}
{d^{(i)}}^2 &=& - \Omega \, \sum_{j=1}^{N} g^{(ij)} \sigma^{(ij)} \nonumber \\
d^{(i)} d^{(j)} - d^{(j)} d^{(i)} &=& \Omega \, \left( ( p_i - p_j ) \, g^{(ij)} + \sum_{k=1}^{N} g^{(jk)} \sigma^{(ik)} -\sum_{k=1}^{N} g^{(ik)} \sigma^{(jk)}  \right) 
\label{eq:4.15a}
\end{eqnarray}
with no implicit summation of any repeated labels. 


\subsection{Massive multi-form gauge theory}

It is evident, for example from ({\ref{eq:4.15a}}), that relations between multi-form operators in curved space become very complicated in the most general case. This makes a full classification of all gauge fields, corresponding to arbitrary Young tableaux, more difficult. Nevertheless, recall that the analysis of the extra compensator fields required to maintain gauge-invariance in the massive system can be done in flat space. We therefore perform this analysis for general multi-form gauge fields that are irreducible under $GL(D,{\mathbb{R}})$.   

Consider a gauge potential that is a tensor $A$ in an arbitrary irreducible representation of $GL(D,{\mathbb{R}})$ whose components have the index symmetry of an $N$-column Young tableau with $p_i$ cells in the $i$th column (it is assumed $p_i \geq p_{i+1}$). A given multi-form $A \in X^{p_1 ,..., p_N}$ is in such an irreducible representation if
%
%
\be
\sigma^{(ij)} A \; =\; 0
\label{eq:4.15}
\ee
for any $j \geq i$ and also satisfying $\sigma^{(ji)} A =0$ only if the $i$th and $j$th columns are of equal length, such that $p_i = p_j$. Such a representation is labeled $[ p_1 ,..., p_N ]$. One can project onto this irreducible tensor subspace $X^{[p_1 ,...,p_N ]}$ from $X^{p_1 ,...,p_N}$ using the Young projector ${\cal{Y}}_{[p_1 ,...,p_N ]}$.

The gauge transformation for $A$ is given by
%
%
\begin{equation}
\delta A \;\; = \;\; {\cal{Y}}_{[ p_1 ,..., p_N ]} \circ \left( \, \sum^N_{i=1}\; d^{(i)} \alpha^{(i)} \, \right)
\label{eq:4.16}
\end{equation}
for any $N$ gauge parameters $\alpha^{(i)} \,\in\, X^{p_1 ,..., p_i -1,..., p_N }$. As for the bi-form case, we now take these gauge parameters to be irreducible, such that $\alpha^{(i)} \, \in\, X^{[ p_1 ,..., p_i -1,..., p_N ]}$. If the $i$th and $j$th columns of the Young tableau corresponding to $A$ have equal lengths (with $p_i = p_j$) then one can express the $\alpha^{(i)}$ and $\alpha^{(j)}$ gauge transformations in terms of a single parameter. The reason for this follows by taking these particular two columns to form a bi-form subspace of the order $N$ multi-form space, then using the arguments given for bi-forms in previous sections. This argument can be continued so that if {\textit{all}} columns are of equal length then the gauge transformation ({\ref{eq:4.16}}) can be described in terms of a single inequivalent parameter.

Consider now adding a mass term to the gauge-invariant action for $A$ in Minkowski space
\footnote{This massless action is discussed in {\cite{keythesis}}, {\cite{keydeMHul1}}, {\cite{keydeMHul2}} and many earlier references {\cite{keyFro}}, {\cite{keylabmor}}, {\cite{keysieg}}, {\cite{keytsul}}, {\cite{keyFraSag}}.}
such that all these symmetries are broken. Begin by assuming that all columns are of different length so that there are the maximum number of $N$ inequivalent gauge parameters that are all broken. One should therefore introduce $N$ compensator fields $B^{(i)} \,\in \, X^{[ p_1 ,..., p_i -1,..., p_N ]}$ each with its own gauge transformation 
%
%
\begin{equation}
\delta B^{(i)} \;\; = \;\; {\cal{Y}}_{[ p_1 ,..., p_i -1 ,..., p_N ]} \circ \left( \, \sum^N_{j=1}\; d^{(j)} \beta^{(ij)} \, \right)
\label{eq:4.17}
\end{equation}
for any $N^2$ gauge parameters $\beta^{(ij)} \,\in\, X^{[ p_1 ,..., p_i -1,..., p_j -1 ,..., p_N ]}$
\footnote{The brackets enclosing labels $i,j =1,...N$ do not imply any symmetrisation here.}
. This would indicate the requirement of $N^2$ more compensator fields $C^{(ij)} \,\in\, X^{[ p_1 ,..., p_i -1,..., p_j -1 ,..., p_N ]}$. However, just as for the bi-form case, the original gauge transformation $\delta A$ vanishes identically if the parameters
%
%
\begin{equation}
\alpha^{(i)} \;\; = \;\; {\cal{Y}}_{[ p_1 ,..., p_i -1 ,..., p_N ]} \circ \left( \, d^{(i)} \zeta^{(ii)} + \sum_{j \neq i} \; d^{(j)} \zeta^{(ij)} \, \right)
\label{eq:4.18}
\end{equation}
for any $N$ parameters $\zeta^{(ii)} \,\in\, X^{[ p_1 ,..., p_i -2,..., p_N ]}$ and $N(N-1)/2$ parameters $\zeta^{(ij)} = - \zeta^{(ji)} \,\in\, X^{[ p_1 ,..., p_i -1,..., p_j -1 ,..., p_N ]}$. Therefore the gauge transformations of $A$ with parameters ({\ref{eq:4.18}}) are not responsible for any symmetry breaking. Consequently, some of the $N^2$ $C^{(ij)}$ are redundant and, in particular, one requires only $N^2 - \left( N + N(N-1)/2 \right) = N(N-1)/2$ more compensator fields, which can be arranged by imposing $C^{(ij)} = - C^{(ji)}$. This procedure continues in a straightforward way. That is, each $C^{(ij)} \,\in \, X^{p_1 ,..., p_i -1,..., p_j -1,..., p_N }$ has its own canonical gauge transformations in terms of the $N^2 (N-1)/2$ parameters $\varepsilon^{(ijk)} = - \varepsilon^{(jik)} \,\in\, X^{[ p_1 ,..., p_i -1,..., p_j -1 ,..., p_k -1,..., p_N ]}$ though not all of these correspond to independent, non-trivial gauge transformations of $A$. In particular, $\delta A$ vanishes identically if the gauge parameters 
%
%
\begin{equation}
\alpha^{(i)} \;\; = \;\; {\cal{Y}}_{[ p_1 ,..., p_i -1 ,..., p_N ]} \circ \left( \, \sum_{j,k =1}^{N} \; d^{(j)} d^{(k)} \zeta^{(ijk)} \, \right)
\label{eq:4.18a}
\end{equation}
for any parameters $\zeta^{(ijk)} = \zeta^{(ikj)} = - \zeta^{(jik)} \,\in\, X^{[ p_1 ,..., p_i -1,..., p_j -1 ,..., p_k -1,..., p_N ]}$. This means that $\zeta^{(ijk)}$ is in the irreducible representation of the symmetric group $S_N$ corresponding to the Young tableau
\, {\tiny \begin{tabular}{|c|c|}\hline
& \\ \hline
   \\ \cline{1-1}
\end{tabular}} \,
in addition to it being in the irreducible tensor representation $[ p_1 ,..., p_i -1,..., p_j -1 ,..., p_k -1,..., p_N ]$ of $GL(D,{\mathbb{R}})$. Therefore the gauge transformations of $A$ with parameters ({\ref{eq:4.18a}}) are not responsible for any symmetry breaking and one needs only $N(N-1)(N-2)/6$ of the compensator fields $D^{(ijk)}$ for the gauge transformations of $C^{(ij)}$, which is achieved by demanding total antisymmetry in the labels, such that $D^{(ijk)} = - D^{(jik)} = - D^{(ikj)}$.
\footnote{This follows by subtracting the $dim ($
{\tiny \begin{tabular}{|c|c|}\hline
& \\ \hline
   \\ \cline{1-1}
\end{tabular}} 
$)$
trivial gauge parameters $\zeta^{(ijk)}$ from the original \\
$dim ($
{\tiny \begin{tabular}{|c|}\hline
 \\ \hline
   \\ \cline{1-1}
\end{tabular}} 
$\otimes$
{\tiny \begin{tabular}{|c|}\hline
   \\ \cline{1-1}
\end{tabular}} 
$)$
parameters $\varepsilon^{(ijk)}$, resulting in $dim ($
{\tiny \begin{tabular}{|c|}\hline
 \\ \hline
 \\ \hline
   \\ \cline{1-1}
\end{tabular}} 
$) = N(N-1)(N-2)/6$
non-trivial, independent gauge transformations of $C^{(ij)}$.  
}

In general, at the $r$th iteration of this procedure one has {\scriptsize $\left( \matrix{N \cr r \cr} \right)$} new independent compensator fields $H_{r}^{( i_1 ... i_r )} \,\in\, X^{[ p_1 ,..., p_{i_1} -1,..., p_{i_r} -1,..., p_N ]}$ which are totally antisymmetric in all $r$ labels
\footnote{In this notation, $H_0 := A$, $H_{1}^{(i)} := B^{(i)}$, $H_{2}^{(ij)} := C^{(ij)}$ and $H_{3}^{(ijk)} := D^{(ijk)}$.}
. After the $N$th iteration one has only one independent field $H_{N}^{( 1 ... N )} \,\in\, X^{[ p_1 -1 ,..., p_N -1 ]}$ and this is where the procedure ends since its gauge transformation is not responsible for any symmetry breaking. Therefore one needs a total of $\sum_{r=0}^{N}${\scriptsize $\left( \matrix{N \cr r \cr} \right)$}$\, = 2^N$ fields $( H_0 ,..., H_{r}^{( i_1 ... i_r )} ,..., H_{N}^{( 1...N )} )$ to describe this massive gauge-invariant system. Following the earlier arguments, if the restriction of unequal column lengths is removed then one will generally require less than this number of fields. For example, if all columns are of equal length then one can describe the massive gauge-invariant system in terms of $N+1 \leq 2^N$ fields -- one at each level of iteration, plus $A$. This agrees with the analysis in {\cite{keyZin3}}, {\cite{keyZin2}} for the case of a totally symmetric tensor (corresponding to what we call type $[1,...,1]$, where all columns are of unit length).

\vspace*{.4in}
{\textbf{Note added}} : While this work was in preparation, the two papers {\cite{keyAlk1}} and {\cite{keyAlk2}} appeared which describe massless gauge-invariant theories for fields of mixed symmetry in (anti-)de Sitter space. In particular, {\cite{keyAlk2}} also constructs massless gauge-invariant actions for fields corresponding to any Young tableau with two columns in anti-de Sitter space. These references consider only (partially) massless fields in curved space and they are described via a first order formalism which is somewhat different to the description given here. Nonetheless, there is clearly some overlap in the content of this paper with that of these works.  
   

\vspace*{.4in}
{\textbf{{\large{Acknowledgments}}}}

I would like to thank C. Hull, P. Kumar and B. Spence for helpful comments and useful discussions. I am especially grateful to C. Hull for a critical reading of the paper.



\end{document}